\documentclass[usenatbib,useAMS]{mn2e}
\usepackage{epsfig,psfig,graphicx,float,color}
\usepackage{subfigure}
\usepackage{mybibdefs}
\usepackage{multirow}
\usepackage{widetext}
 
 \usepackage{color}

\newcommand{\Mpc}{\mbox{Mpc}}
\newcommand{\kpc}{\mbox{kpc}}

\newcommand{\km}{\mbox{km}}
\newcommand{\s}{\mbox{s}}
\newcommand{\h}{h}

\usepackage{graphicx}
\usepackage{amssymb}
\usepackage{epstopdf}
\title[Galaxy Zoo: Bar Lengths in Nearby Disk Galaxies]{Galaxy Zoo: Bar Lengths in Local Disk Galaxies$^*$ }  
\author[Hoyle et al.]{Ben  Hoyle$^{1,2,3}$,
   Karen. L. Masters$^2$, Robert C. Nichol$^2$, Edward M. Edmondson$^2$,  \newauthor Arfon M. Smith$^4$, Chris Lintott$^{4,5}$, Ryan Scranton$^6$, Steven Bamford$^{7}$, \newauthor Kevin Schawinski$^{8,9}$, Daniel Thomas$^2$. \\\\\\\\$^1$Institute for Sciences of the Cosmos (ICC-UB, IEEC), University of Barcelona, Marti i Franques 1, Barcelona, 08024 Spain.\\$^2$Institute of Cosmology \& Gravitation,
  University of Portsmouth, Dennis Sciama Building, Portsmouth, PO1.
  3FX, UK.\\$^{3}$Consejo Superior de Investigaciones Cientificas, Serrano 117, Madrid,
28006, Spain.\\$^{4}$Oxford Astrophysics, Department of Physics, University of Oxford, Denys Wilkinson Building, Keble Road, Oxford, OX1 3RH, UK.\\$^5$Adler Planetarium, 1300 S. Lakeshore Drive, Chicago, Illinois, 60605, USA.\\$^6$University of California-Davis, One Shields Avenue, CA 95616.\\$^{7}$Centre for Astronomy \& Particle Theory, University of Nottingham, University Park, Nottingham, NG7 2RD, UK.\\$^8$Einstein Fellow, Department of Physics, Yale University, New Haven, CT 06511, U.S.A.\\$^9$Yale Center for Astronomy and Astrophysics, Yale University, P.O. Box 208121, New Haven, CT 06520, U.S.A.\\\\\\$^*$This publication has been made possible by the participation of more than 200,000 volunteers in the Galaxy Zoo project. \\ Their contributions are individually acknowledged at \texttt{http://www.galaxyzoo.org/Volunteers.aspx}. \\
\\
{\tt E-mail: benhoyle1212@icc.ub.edu}
 }
  
\begin{document}
\date{Accepted ----. Received ----; in original form ----.}
\pagerange{\pageref{firstpage}--\pageref{lastpage}} \pubyear{2010}
\maketitle
\label{firstpage}
\begin{abstract}
We present an analysis of bar length measurements of $3150$ local galaxies in a volume limited sample of low redshift ($z < 0.06$) disk galaxies. Barred galaxies were initially selected from the Galaxy Zoo 2 project, and the lengths and widths of the bars were manually drawn by members of the Galaxy Zoo community using a Google Maps interface. Bars were measured independently by different observers, multiple times per galaxy ($\ge3$), and we find that observers were able to reproduce their own bar lengths to $3$\% and each others' to better than $20$\%. We find a Òcolor bimodalityÓ in our disk galaxy population with bar length, i.e., longer bars inhabit redder disk galaxies and the bars themselves are redder, and that the bluest galaxies host the smallest galactic bars ($< 5 \,\h^{-1}\,\kpc$). We also find that bar and disk colors are clearly correlated, and for galaxies with small bars, the disk is, on average, redder than the bar colors, while for longer bars the bar then  itself is redder, on average, than the disk. We further find that galaxies with a prominent bulge are more likely to host longer bars than those without bulges. We categorise our galaxy populations by how the bar and/or ring are connected to the spiral arms. We find that galaxies whose bars are directly connected to the spiral arms are preferentially bluer and that these galaxies host typically shorter bars. Within the scatter, we find that stronger bars are found in galaxies which host a ring (and only a ring). The bar length and width measurements used herein are made publicly available for others to use ({\tt http://data.galaxyzoo.org}).
\end{abstract}

\section{introduction}
Galactic bars are extended linear structures crossing the center of a substantial fraction of disk galaxies. They are comprised of over densities of both luminous and dark matter \citep[see e.g.][]{2005ARA&A..43..581S} and unlike spiral arms they are significant material asymmetries. Bars therefore are able to contribute to the redistribution of matter in the galaxy through exchanging angular momentum with the spiral arms, disk, bulge and rings \citep[see][]{2004ARA&A..42..603K,Athanassoula:2009dd}. Bars are thought to form by disk instabilities and have been found to occur in up to two thirds of all disk galaxies. Models indicate they can be short-lived, long-lived or periodic, depending on the disk and bulge properties, and the interaction history of the galaxy \citep[see e.g.][]{1985MNRAS.217..127S,Athanassoula:2009dd,2010arXiv1007.2979C}.

The actual fraction of disk galaxies that contain bars has been shown to depend on the method of bar detection and the sample selection, for example, \cite{Eskridge:1999pn} used near infrared images of $186$ spiral galaxies, and found that $\sim 70\%$ of galaxies were barred when visual inspected, and \cite{MenendezDelmestre:2006en} using $151$ $2$MASS \citep[][]{Jarrett:2003uy} galaxies, found $\sim60\%$ contained bars, as identified by fitting ellipses to the light profiles. Furthermore,  \cite{Giordano:2010gg} used $253$ disk galaxies in the Virgo cluster and found that $\sim30\%$ were barred after visual inspection of high resolution near infrared images. Recently, \cite{Nair:2010xh} found that the fraction of barred spiral galaxies is a strong function of stellar mass and star formation history, and \cite{Masters:2010rw} studied the fraction or spiral galaxies with bars as a function of galaxy properties, and found that the fraction increases for redder galaxies and those galaxies hosting a bulge.
 
Bar dimensions are equally difficult to define and measure, but techniques such as visual estimation, ellipse fitting, or methods based on Fourier analysis have been used \citep[e.g.][]{1985ApJ...288..438E}. Where measurements have been made, bars have been found to be longer by a factor of $\sim 2.5$ in early-type disk galaxies than late-type disk galaxies \citep[][]{Erwin:2005si}.  

These early observational results have been complemented by N-body simulations, but due to the computational difficulty involved in producing barred galaxies in full hydrodynamic simulations, recent works have only been able to examine the properties of a handful of barred systems, and their results, while providing insight for the creation and roles of bars in galaxies, have yet to be tested on large observational data-sets \cite[e.g. see][]{Athanassoula:2009dd}. The premise of this paper is to present the results of analysis on a large sample of observationally identified barred galaxies, and compare with available simulations.

Early work by \cite{Athanassoula:2003wd} combined theoretical descriptions of bar, disk and bulge components and a suite of $160$ different $\sim 10^{6}$ dark matter particle simulations. She finds that bars are stronger, and rotate slower, in the presence of a large bulge, which is attributed to an exchange of angular momentum, in agreement with an earlier observational studies of $32$ galaxies \citep{G1980A} and more recently with $300$ galaxies \citep{Gadotti:2010ct}. Further observational work described in \cite{2004ARA&A..42..603K} combine results from theory and simulations, which suggests bars drive gas inwards, building a bulge which may then play a role in diminishing the bar \citep[see also][]{kormendy1975,1985ApJ...288..438E}.

\cite{Scannapieco:2010nw} simulate $8$ isolated Milky Way mass galaxies at redshift $z=0$ \citep[see also][for details]{Scannapieco:2008cm} using $\sim 10^6$ particles, and include prescriptions for gas, star formation, chemical enrichment, cooling and feedback, and find that bar, disk and bulge colors are correlated, and that simulated bars are predominantly found in bluer galaxies. The authors go to great care to present their results such that they can be directly compared with Sloan Digital Sky Survey \citep[][hereafter SDSS]{SDSSDR7} observations such as those exhibited in this paper. 

Galactic bars have been observed to be connected in different ways to the spiral arms (if present), or to a ring structure encompassing the bar (if present) - in which case the bar and spiral arms are not necessarily aligned  \citep[for a review of bar galaxy morphology, see][]{2005ARA&A..43..581S}.  The presence of a ring was found to be more likely in strongly barred  systems by \cite{2004ARA&A..42..603K}, although observations of $147$ galaxies \citep[][]{Buta:2005ym}, suggest that, in contrast to the above result, galaxies with stronger bars have spiral arms, and that weaker bars are more likely to be ringed, when compared with the global average. 

In this paper, we attempt to test further the above simulations using hand drawn bar length measurements by multiple observers and visual classifications of $3150$ SDSS galaxies, by asking the following questions;
\begin{itemize}
\item{How are the colors of galaxies, bars and disks correlated? }
Simulations suggest bar and disk colors are correlated  \citep{Scannapieco:2010nw}.
\item{How do the colors change as a function of bar length?}
Simulations show that longer bars inhabit bluer disk galaxies \citep{Scannapieco:2010nw}.
\item{Are other galaxy properties affected by bar length?}
Simulations find that galaxies with a large central bulge host longer bars  \citep{Athanassoula:2003wd}.
\item{Is bar strength or length correlated with the presence of a ring?}
Previous observations suggest that galaxies with a ring have weaker bars \citep{Buta:2005ym} than the population average.

Additionally to the above questions, we also aim to understand;
\item{How is the bar-to-spiral arm connection different in galaxies with longer bars?}
\end{itemize}

To achieve these aims, we initiated a satellite Galaxy Zoo project using data drawn from interim results of  Galaxy Zoo 2\footnote{http://zoo2.galaxyzoo.org} \citep[hereafter GZ2, see Lintott et al. in prep.][]{Masters:2010rw}, which itself is an extension of the original Galaxy Zoo project \citep[][]{Lintott:2008ne,Lintott:2010bx}. We examine the bar and galaxy properties of $3150$ galaxies using a Google Maps powered website which allows members of the Galaxy Zoo community to measure the lengths and widths of bars in disk galaxies selected from GZ2. This provides bar measurements of $3150$ barred galaxies. We also collect data describing the connection of the spiral arms (if they exist) to the ring (if it exists) and the bar.

\cite{Masters:2010rw} analysed $13,665$ galaxies imaged by the  SDSS  and visually identified as disk galaxies from GZ2, and observed that the fraction of barred galaxies increases from $10\%$ to $50\%$ with the prominence of the central bulge \citep[or {\tt fracdeV}\footnote{The fraction of the best fit light profile which comes from a de Vaucouleurs fit as opposed to an exponential fit to the SDSS r-band light profile.}, see][]{Masters:2010hm}. They also find that redder ($g-r$) ``early-type" spiral galaxies are more likely to host galactic bars than their bluer ``late-type" spiral counterparts. We loosely make the distinction between early and late-type disk galaxies using the color cut  $g-r = 0.6$ \citep{2010MNRAS.405..783M}.

The format of the paper is the following, in Section \ref{website} we describe the bar drawing website and present usage statistics.  We describe the input data sample, and observer statistics and agreement in Section \ref{data}. We continue in Section \ref{bar_disk_prop} by deriving bar properties and measuring ``bar"  and ``disk" colors and correlating bar properties against galaxy properties, and then show statistics split by how the spiral arms are connected to the bar.  We discuss implications and extensions to this work and conclude in Section \ref{conclus}. To calculate distances, we assume a flat $\Lambda$CDM cosmology with $\Omega_m, \Omega_{\Lambda},H_0=(0.3,\,0.7,\,70 \,\km\, \s^{-1}\,\Mpc^{-1})$ and $h=H_0/100$.

 \begin{table}
\begin{center}
  \begin{tabular}{l l } 
Acronym & Description  \\ \hline
NSR & Spiral arms and ring are not present. \\ \hline
\multirow{2}{*}{OR} &  Spiral arms are not present \\
& and the bar is connected to the ring. \\ \hline
\multirow{2}{*}{R}&Spiral arms are connected to \\
& the ring around the bar. \\ \hline
S &Spiral arms are connected to the end of the bar. \\ \hline
\multirow{2}{*}{SR} & Spiral arms are connected\\
& to  a mixture of a ring and the bar. \\ \hline
U& The observer is unsure or is unable to decide. \\  \hline
  \end{tabular}
\caption{\label{bar_spirl_table} Possible configurations for connecting the spiral arms (if present) to the ring (if present) and the bar in barred galaxies. The observers were asked to select the most suitable category.}
\end{center}
\end{table}

\section{The Bar Drawing Website}
\label{website}
The bar drawing website\footnote{http://www.icg.port.ac.uk/$\sim$hoyleb/bars/ shows a working example, but is not collecting data.} uses HTML and Javascript to call the Google maps API\footnote{http://code.google.com/apis/maps/} interface with the parameter {\tt mapTypes} set to  {\tt G\_SKY\_MAP\_TYPES} to view celestial maps. Composite color images ($g$, $r$ and $i$ band) from SDSS Data Release 6 \cite[DR6,][]{2008ApJS..175..297A} (excluding the southern stripes) are available in the Google maps interface and we center the map on the galaxy to be classified, and provide a galaxy marker if the observer was unsure of which galaxy to classify. Some Google Maps functionality is available to the observer, e.g., the adjustment of the level of zoom, and the examination of nearby space through dragging of the map. 

The SDSS images in Google Maps have varying levels of quality, some of which are lower than the original SDSS images. The varying quality allows galaxies to be viewed to different zoom levels in the Google Maps interface. We choose a standard zoom level suitable for most galaxies, but some galaxies appeared as a blank screen, and the zoom adjustment allowed these galaxies to be viewed, additionally, it was possible to magnify some galaxies. The variable quality of the images means that images cannot be magnified to the same detail as is available in, for example, the SDSS tool {\tt Navigate}\footnote{http://cas.sdss.org/dr7/en/tools/chart/navi.asp}, and we therefore apply a maximum redshift cut on the galaxy sample. The power of Google Maps comes from the ease with which the included Javascript libraries enable polygons to be drawn on the galaxy images (e.g. to trace the bars), and the polygon properties to be recorded.

\begin{figure*}
   \centering \large{NSR}
   \includegraphics[scale=0.285]{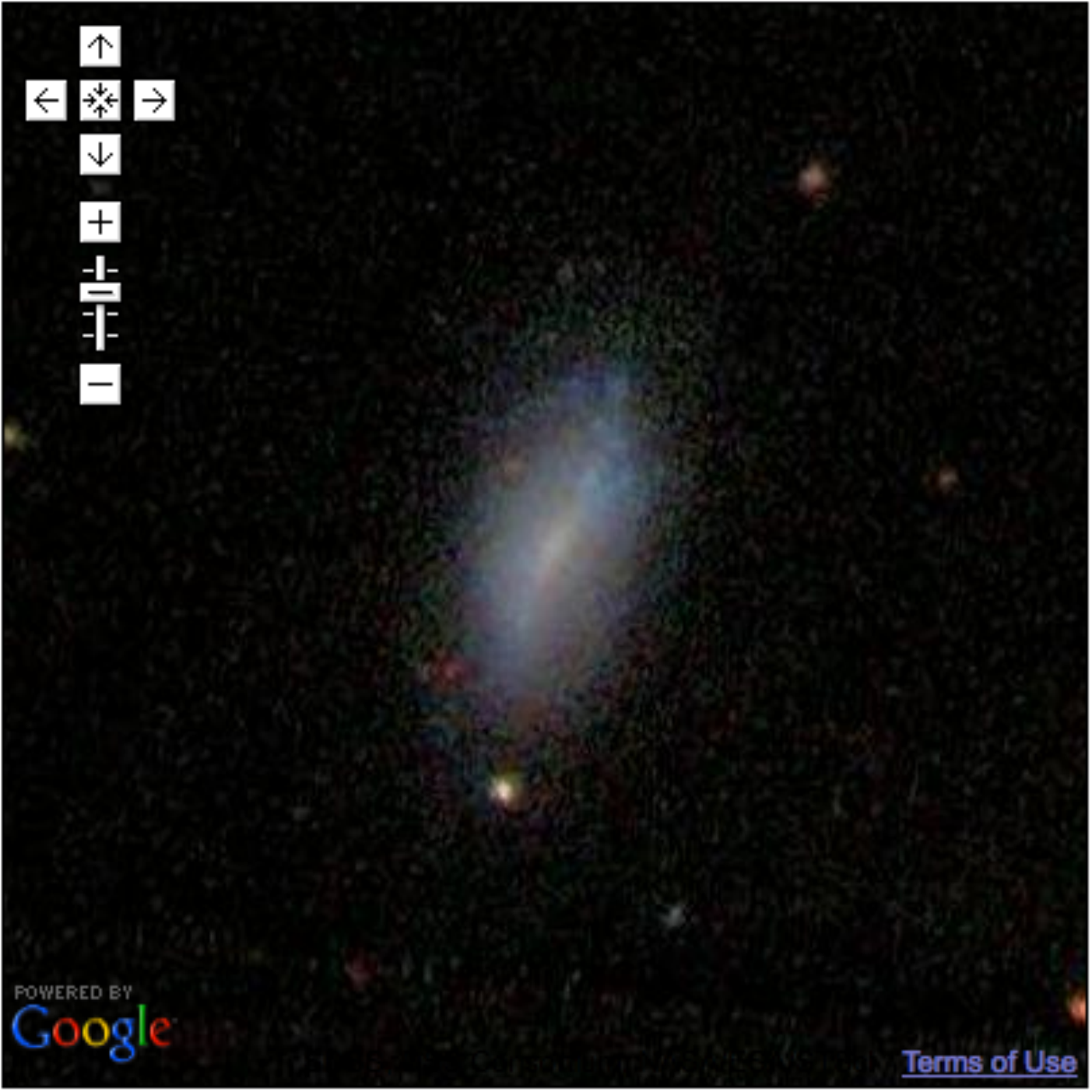}
   \includegraphics[scale=0.285]{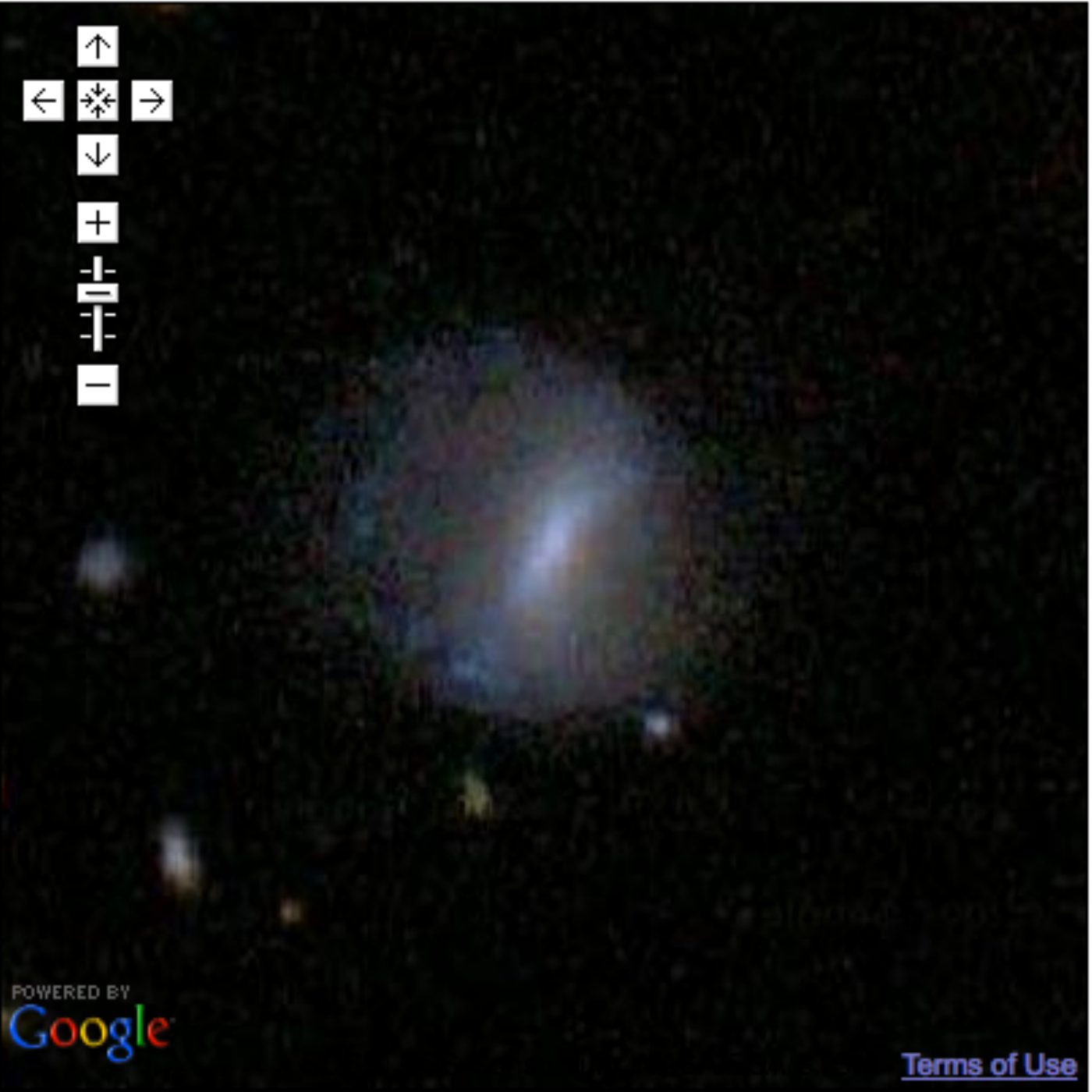}
      \includegraphics[scale=0.285]{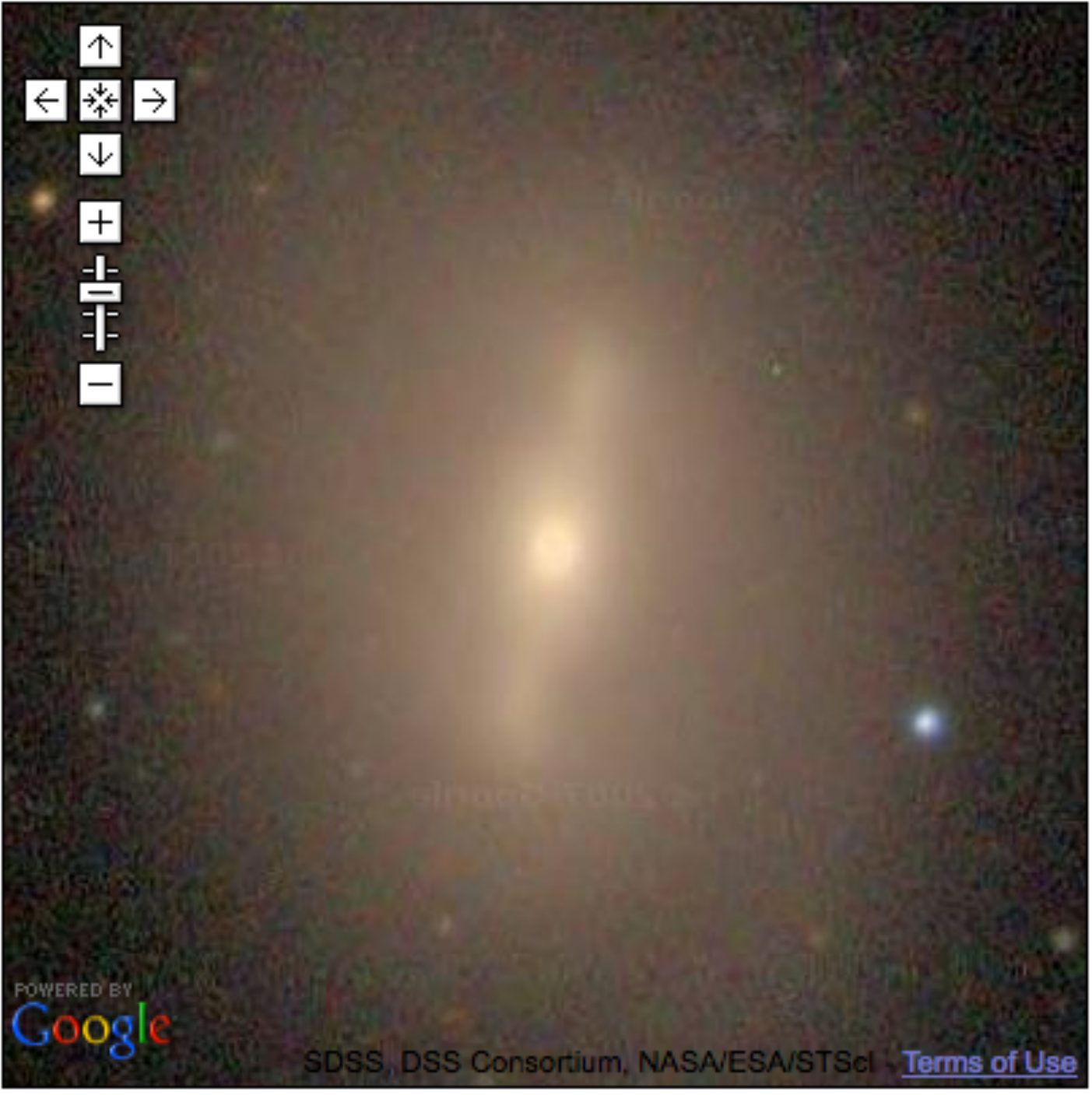}
         \includegraphics[scale=0.285]{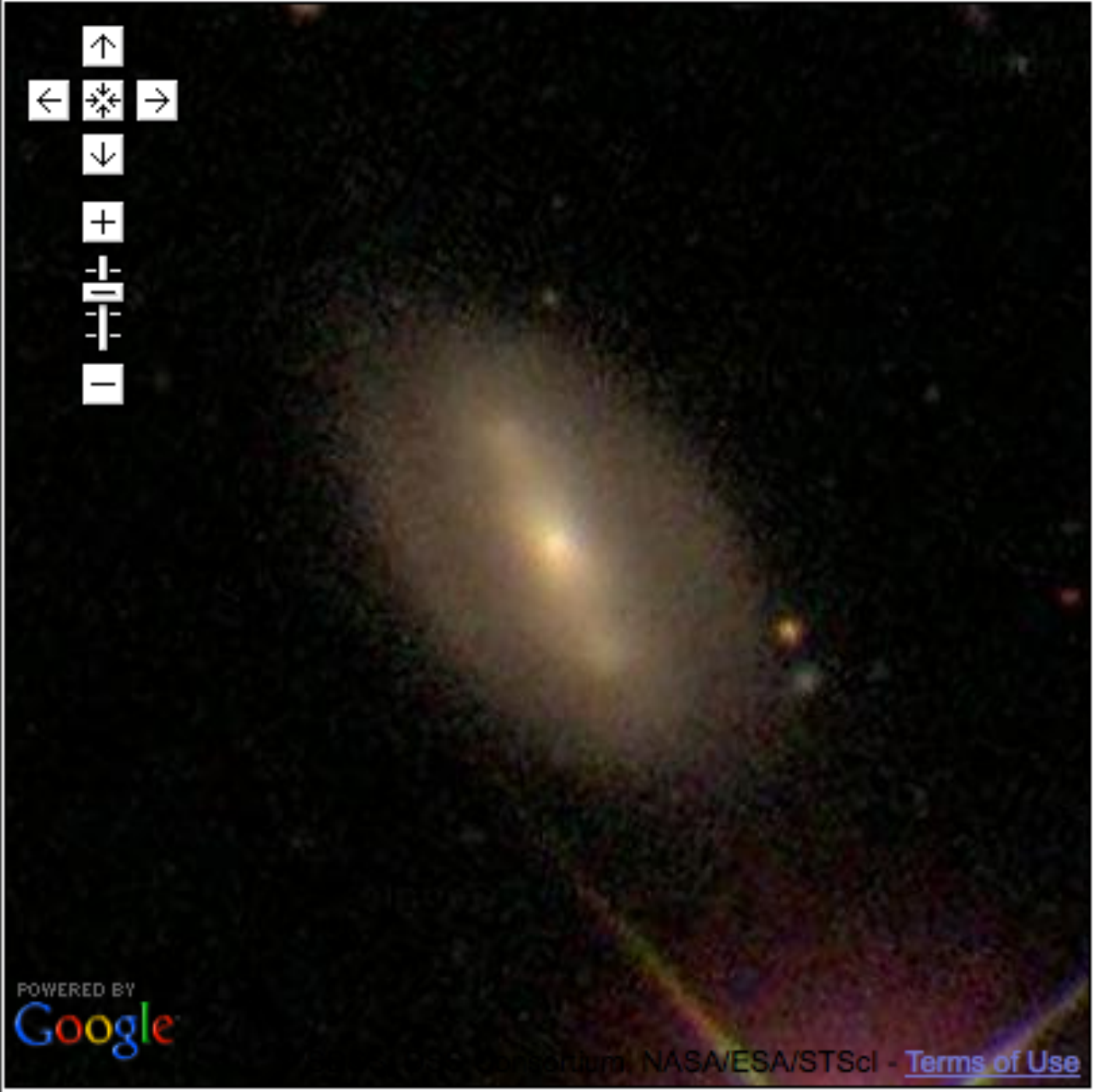} \\
             \large{OR $\,$}
            \includegraphics[scale=0.285]{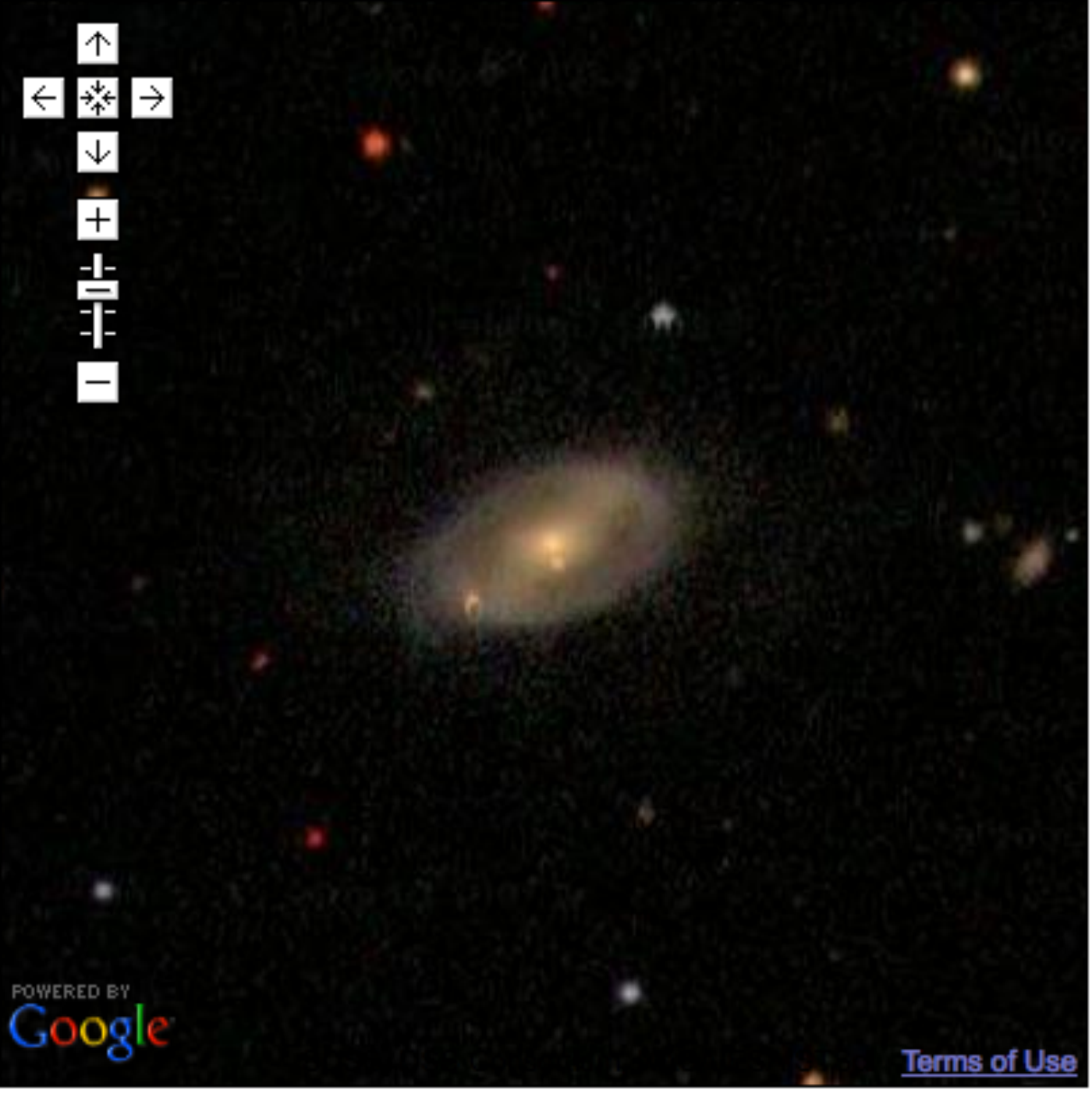}
   \includegraphics[scale=0.285]{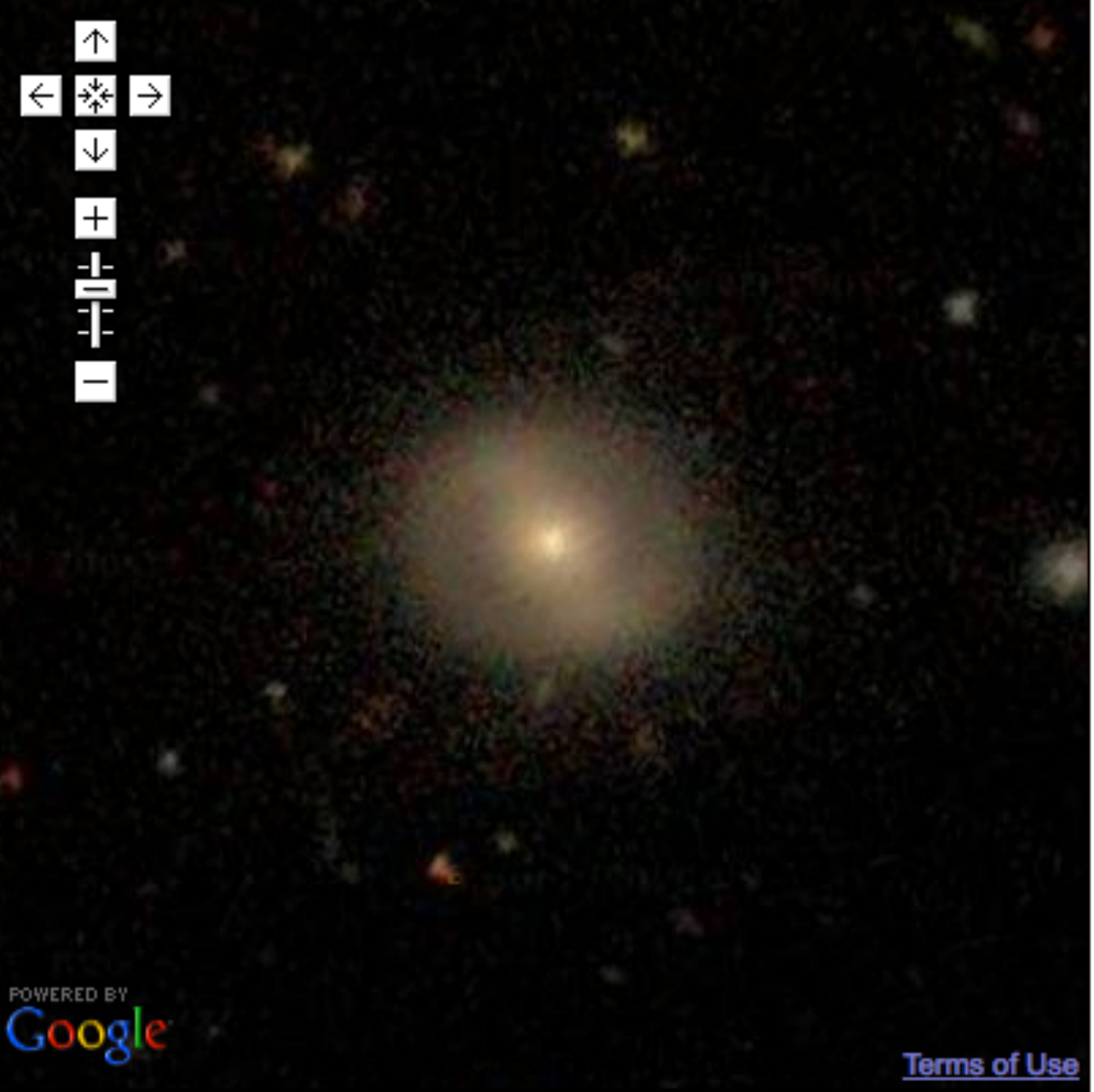}
      \includegraphics[scale=0.285]{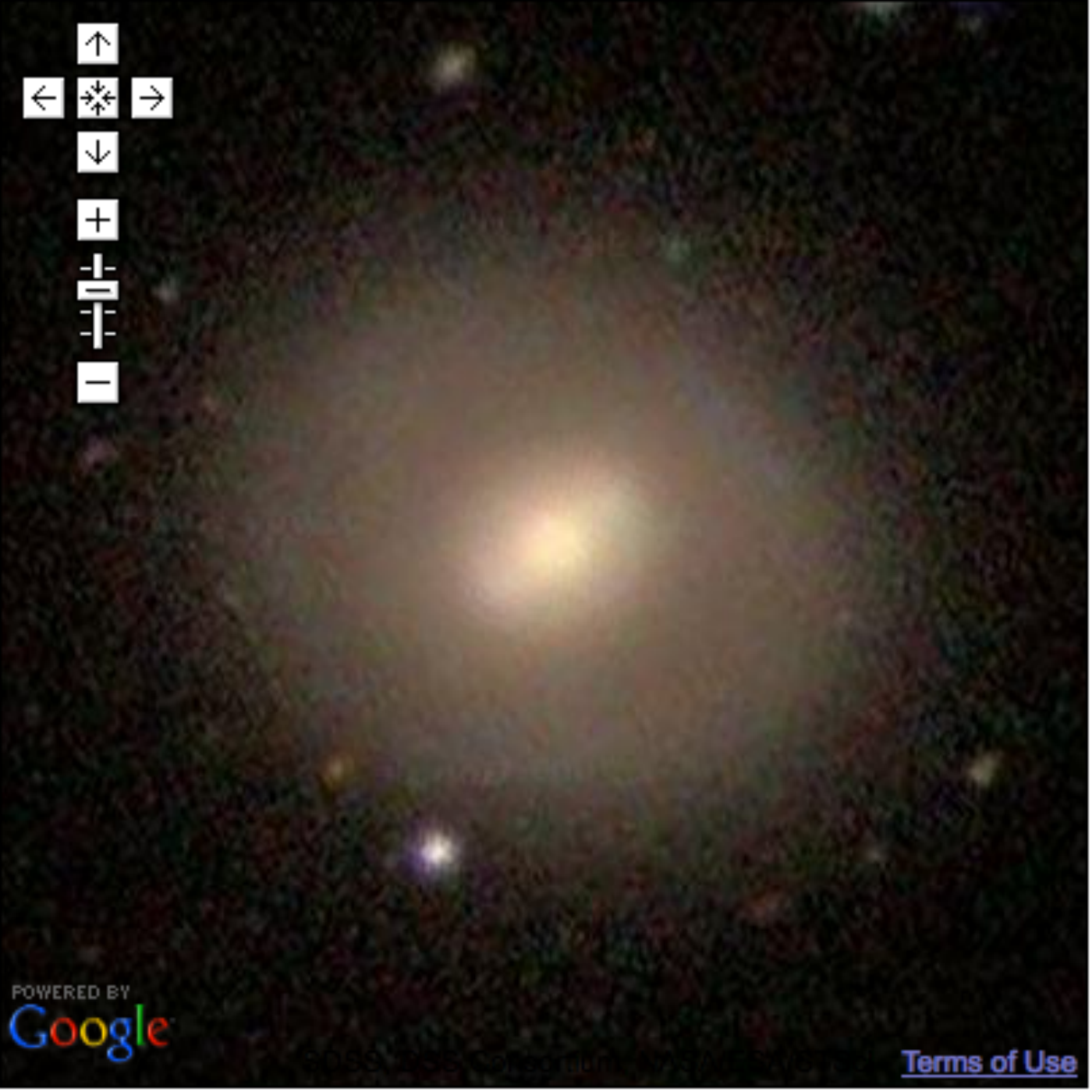}
         \includegraphics[scale=0.285]{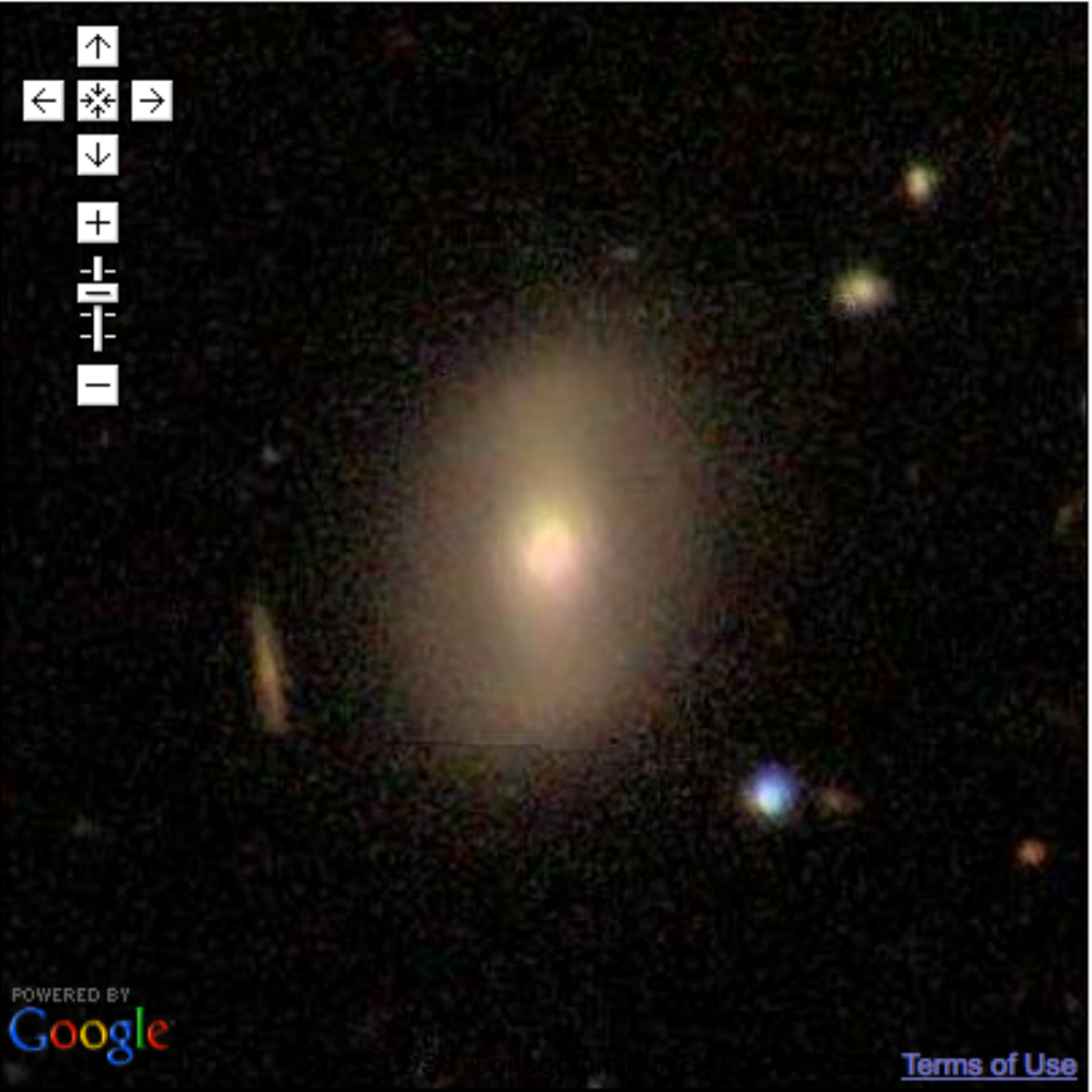} \\
           \large{R $\;$ $\,$}
            \includegraphics[scale=0.285]{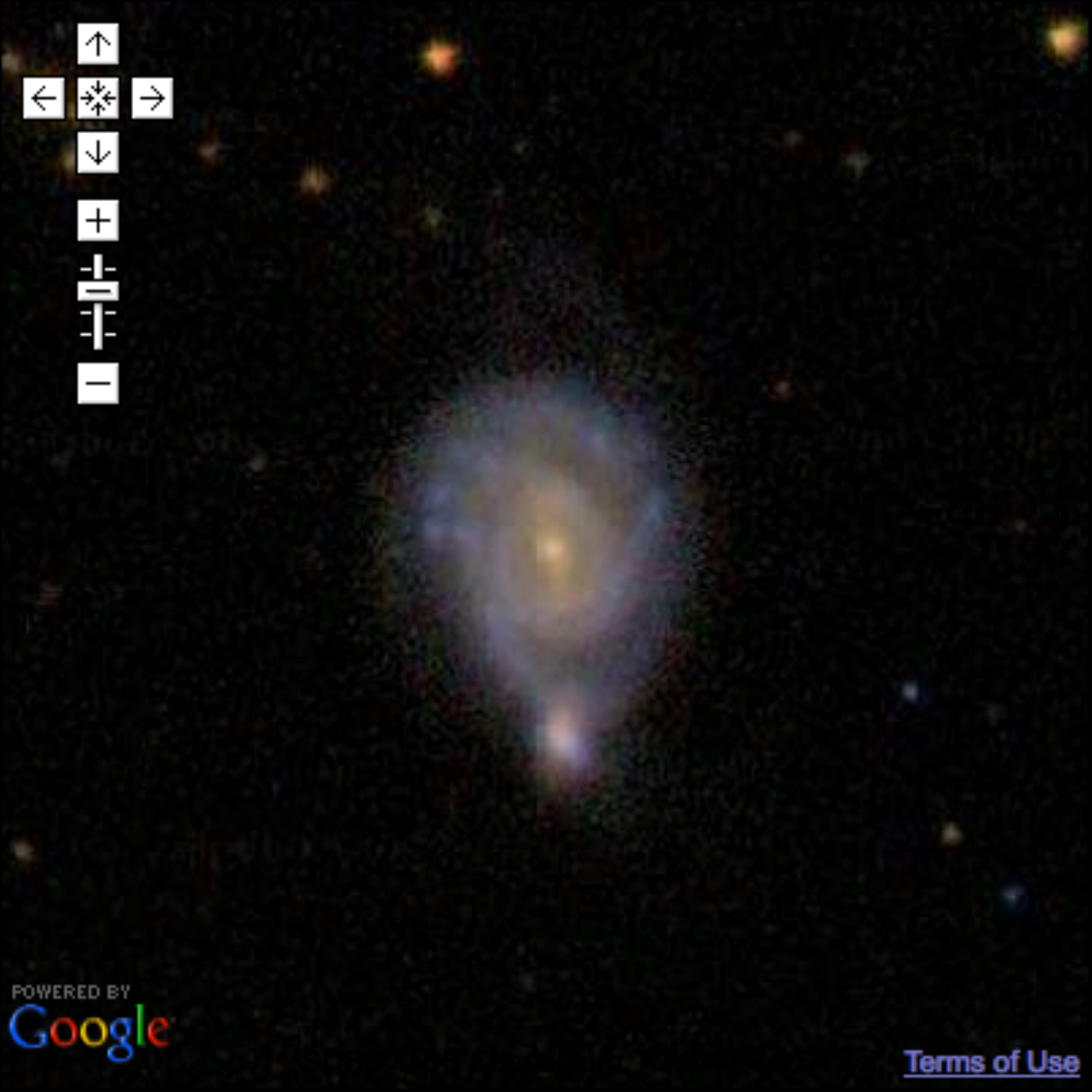}
   \includegraphics[scale=0.285]{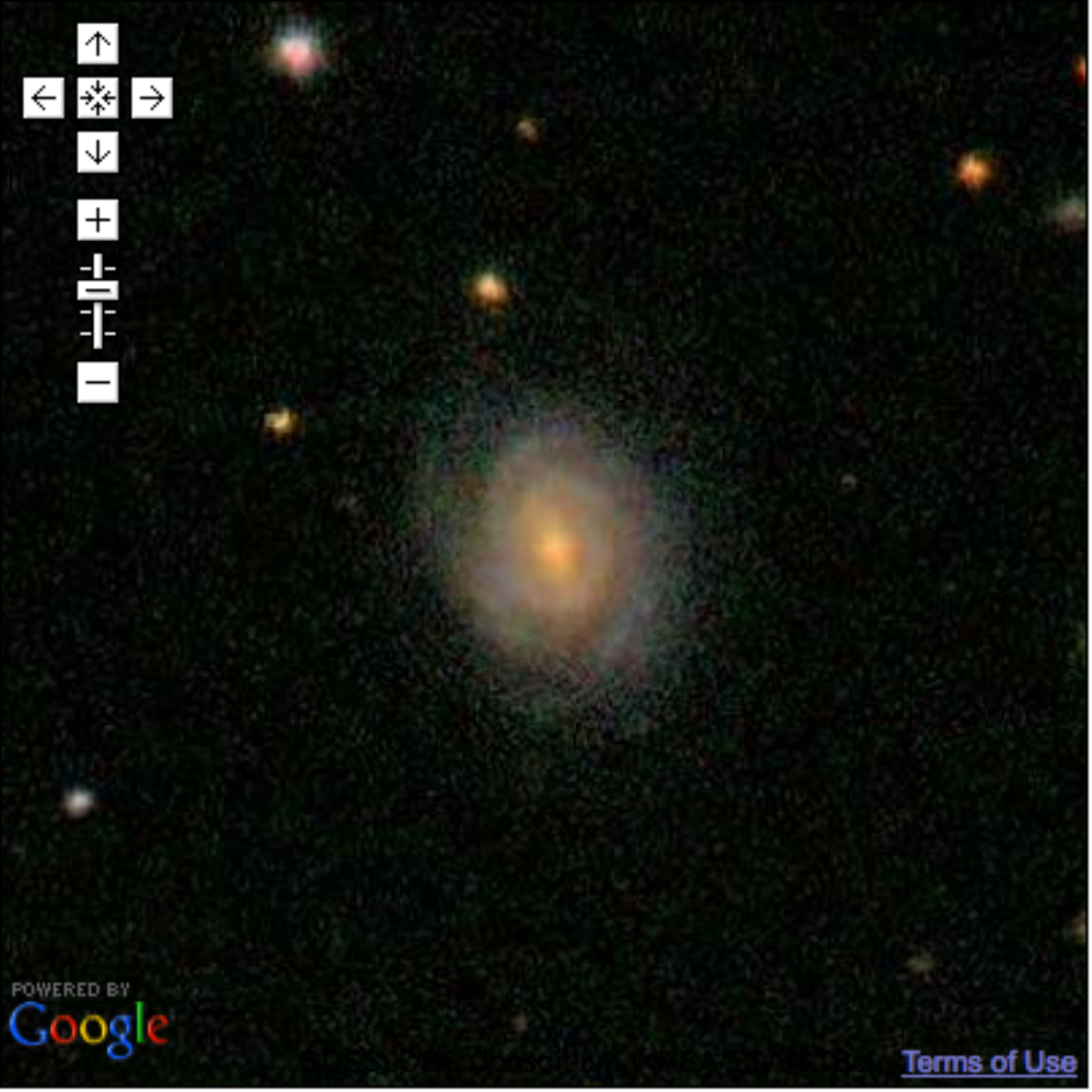}
      \includegraphics[scale=0.285]{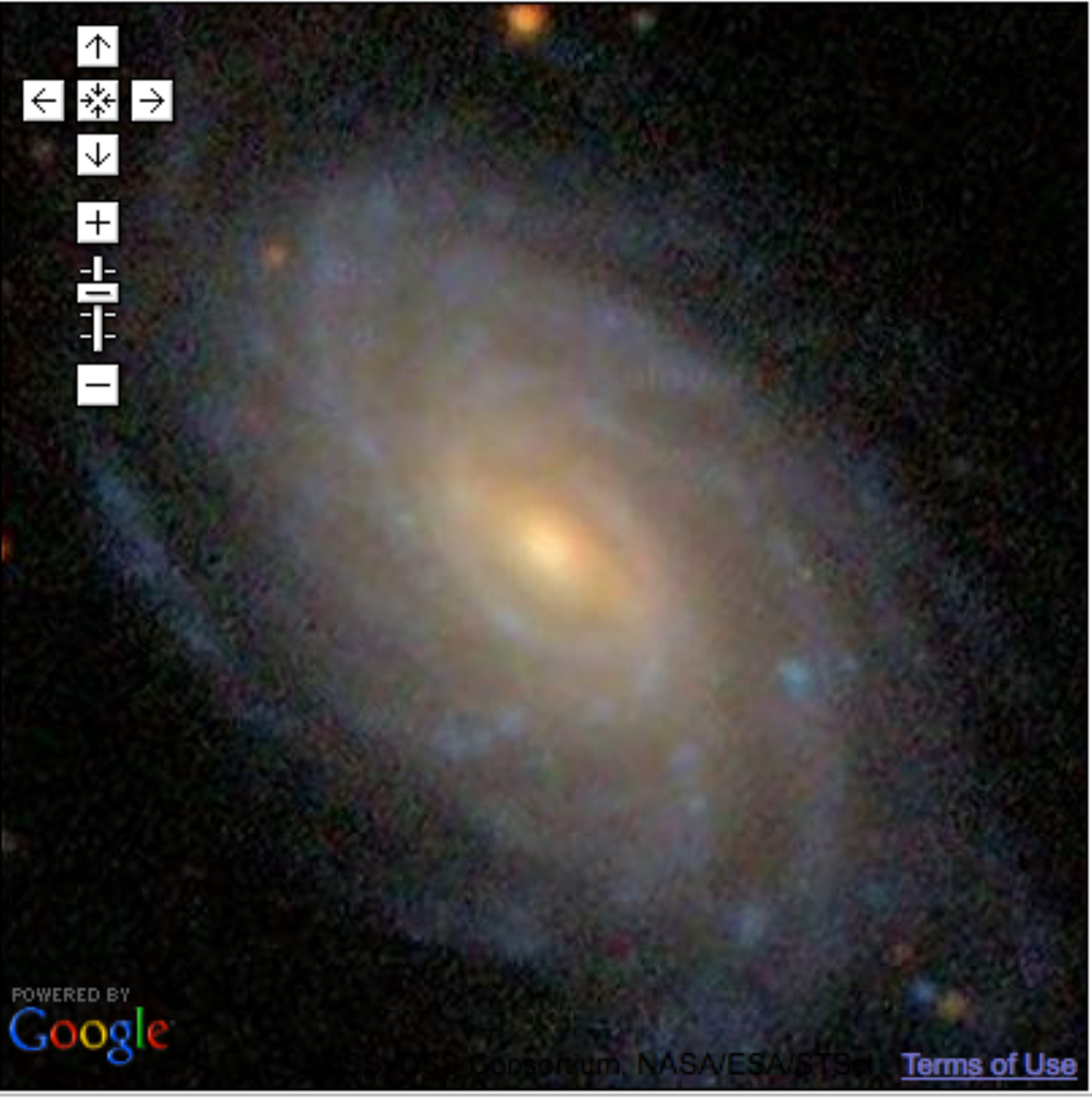}
         \includegraphics[scale=0.285]{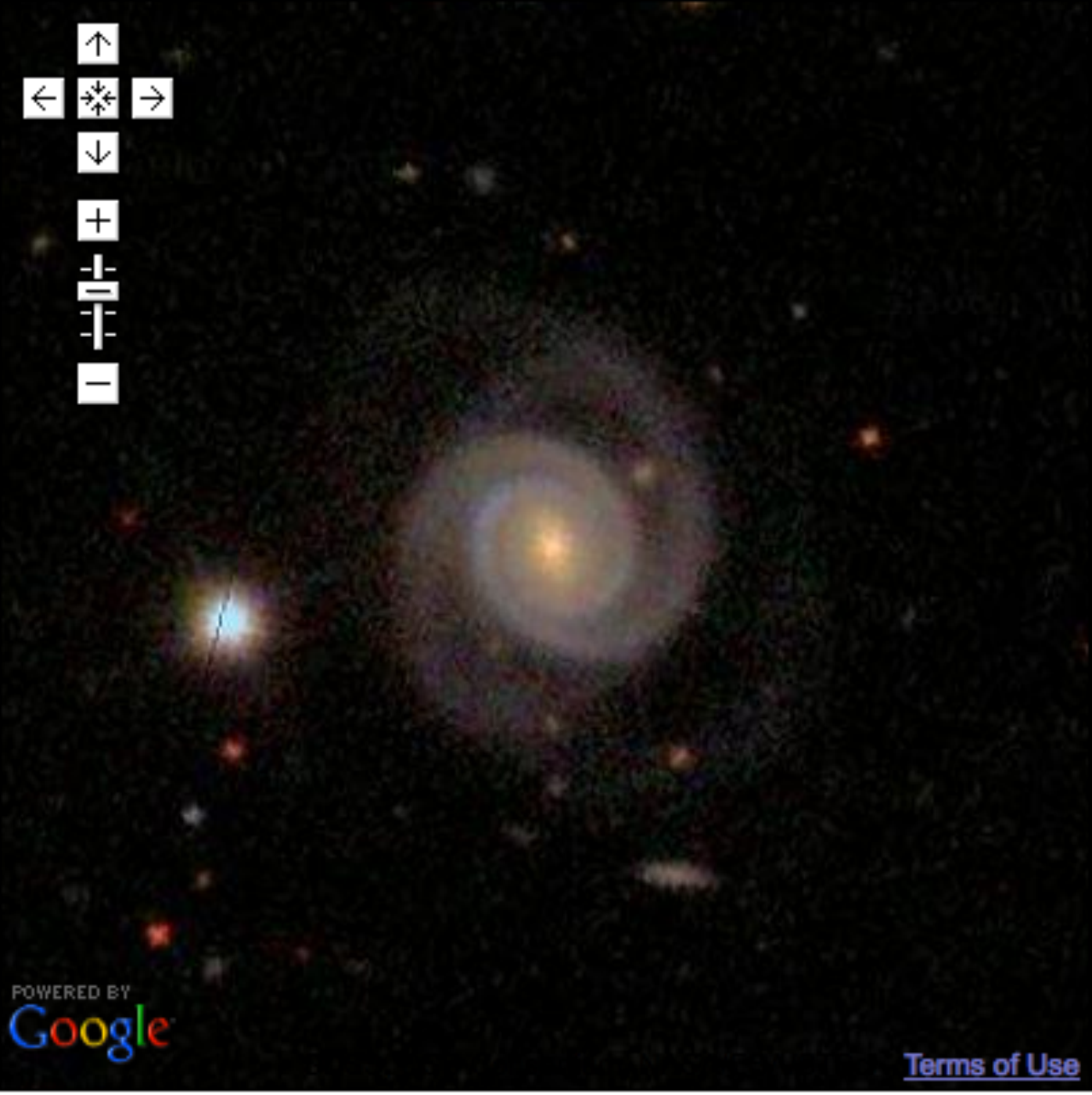} \\
           \large{S $\;$  $\:$}
            \includegraphics[scale=0.285]{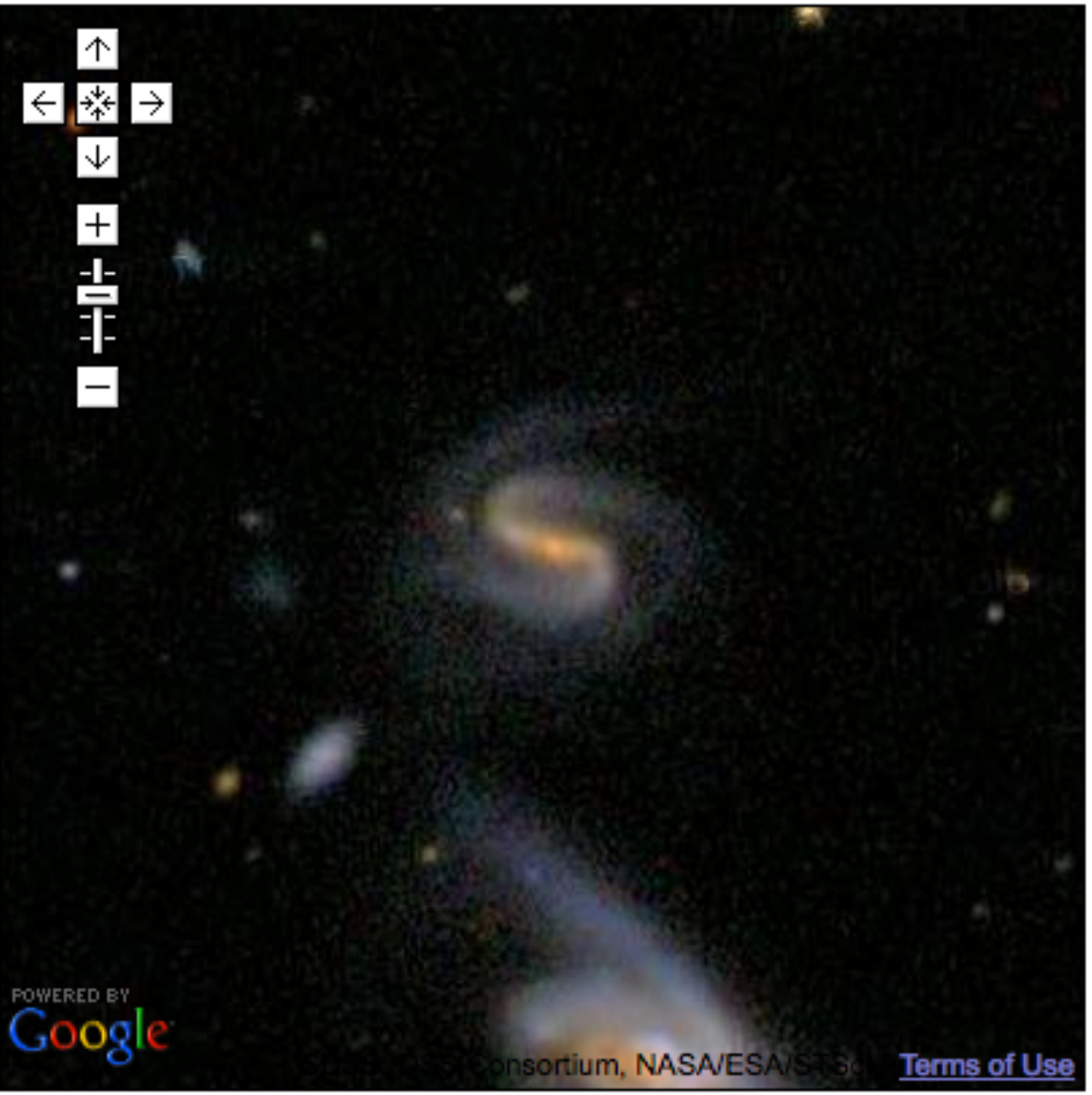}
   \includegraphics[scale=0.285]{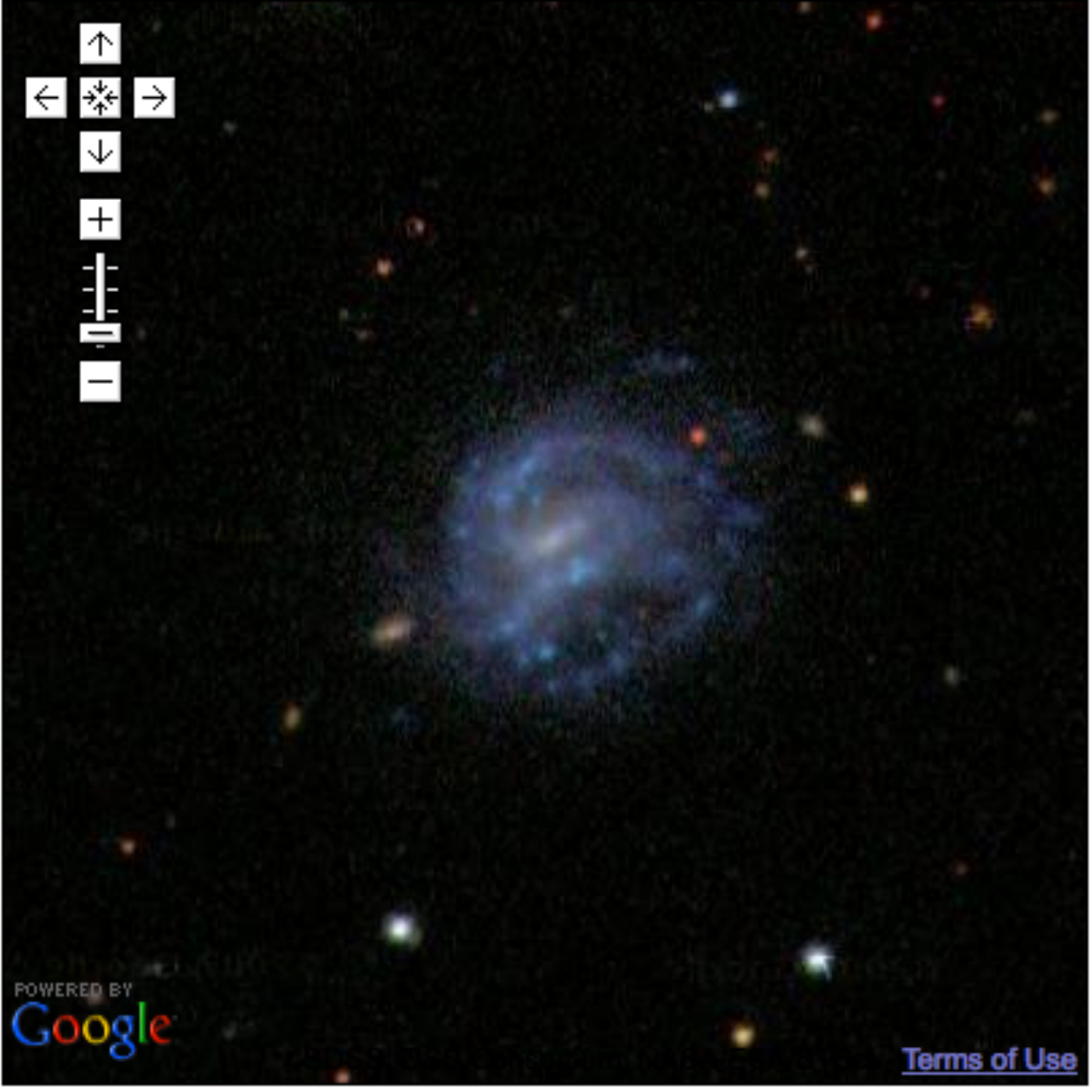}
      \includegraphics[scale=0.285]{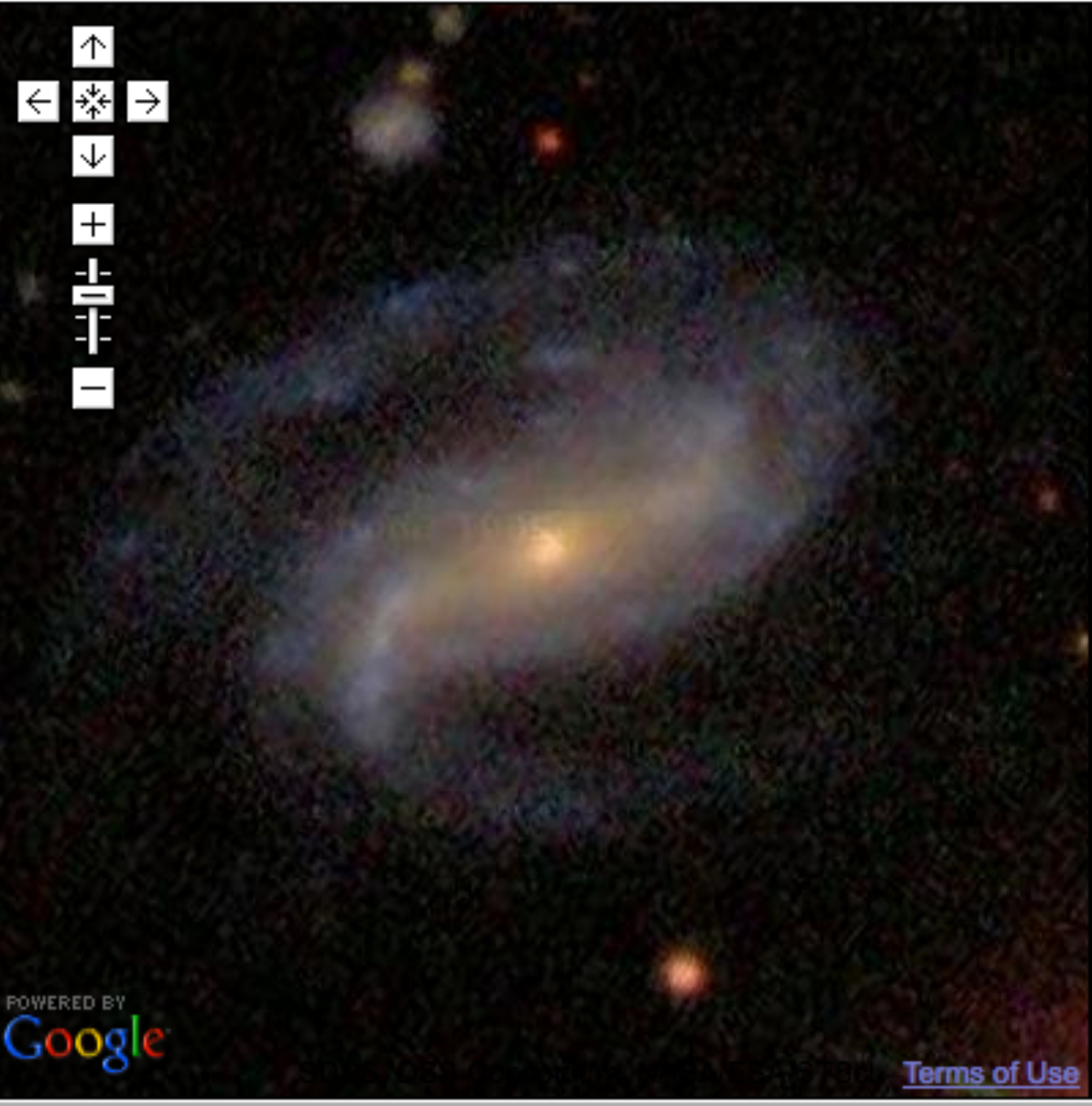}
         \includegraphics[scale=0.285]{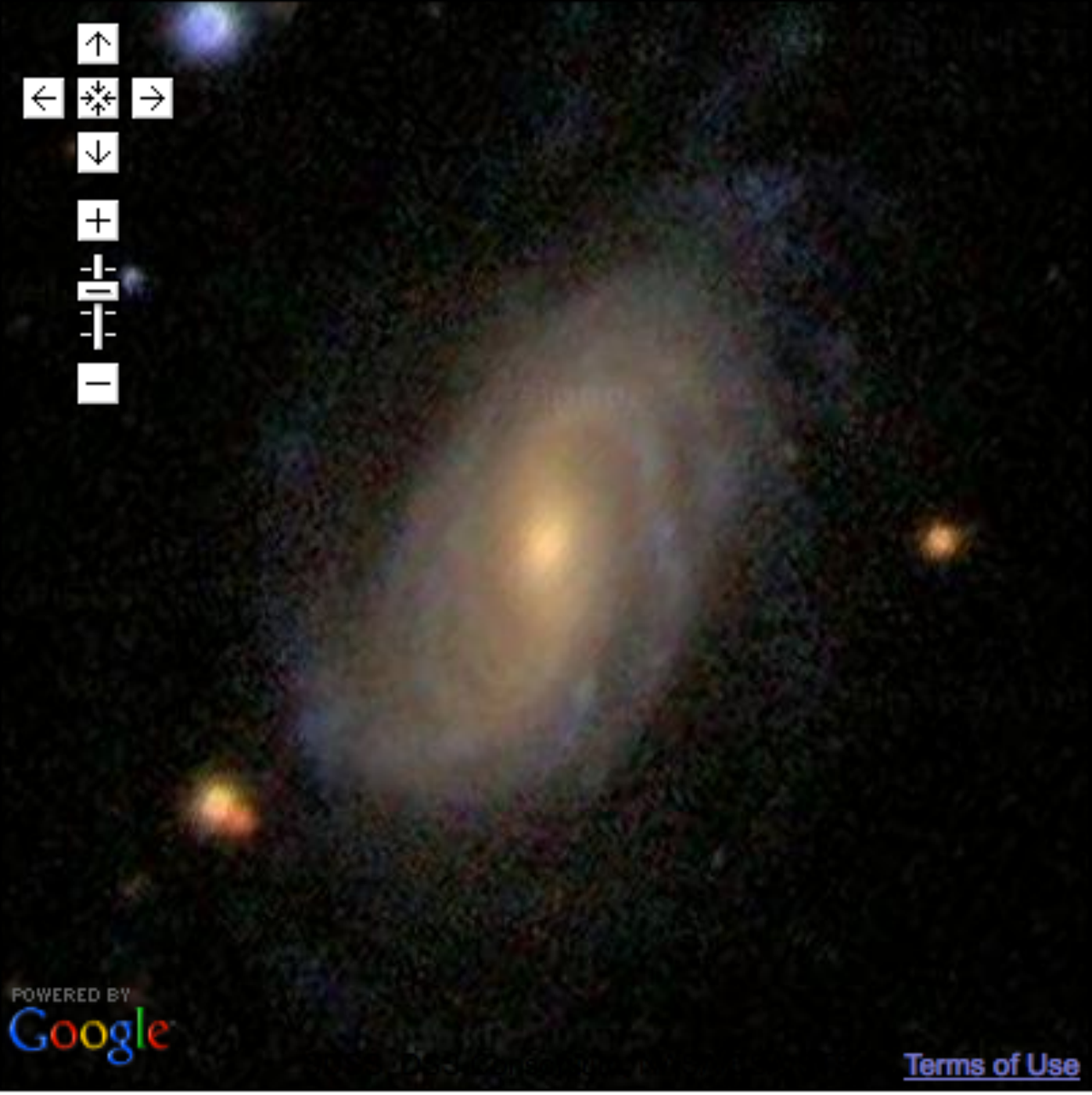} \\
           \large{SR }
           \includegraphics[scale=0.285]{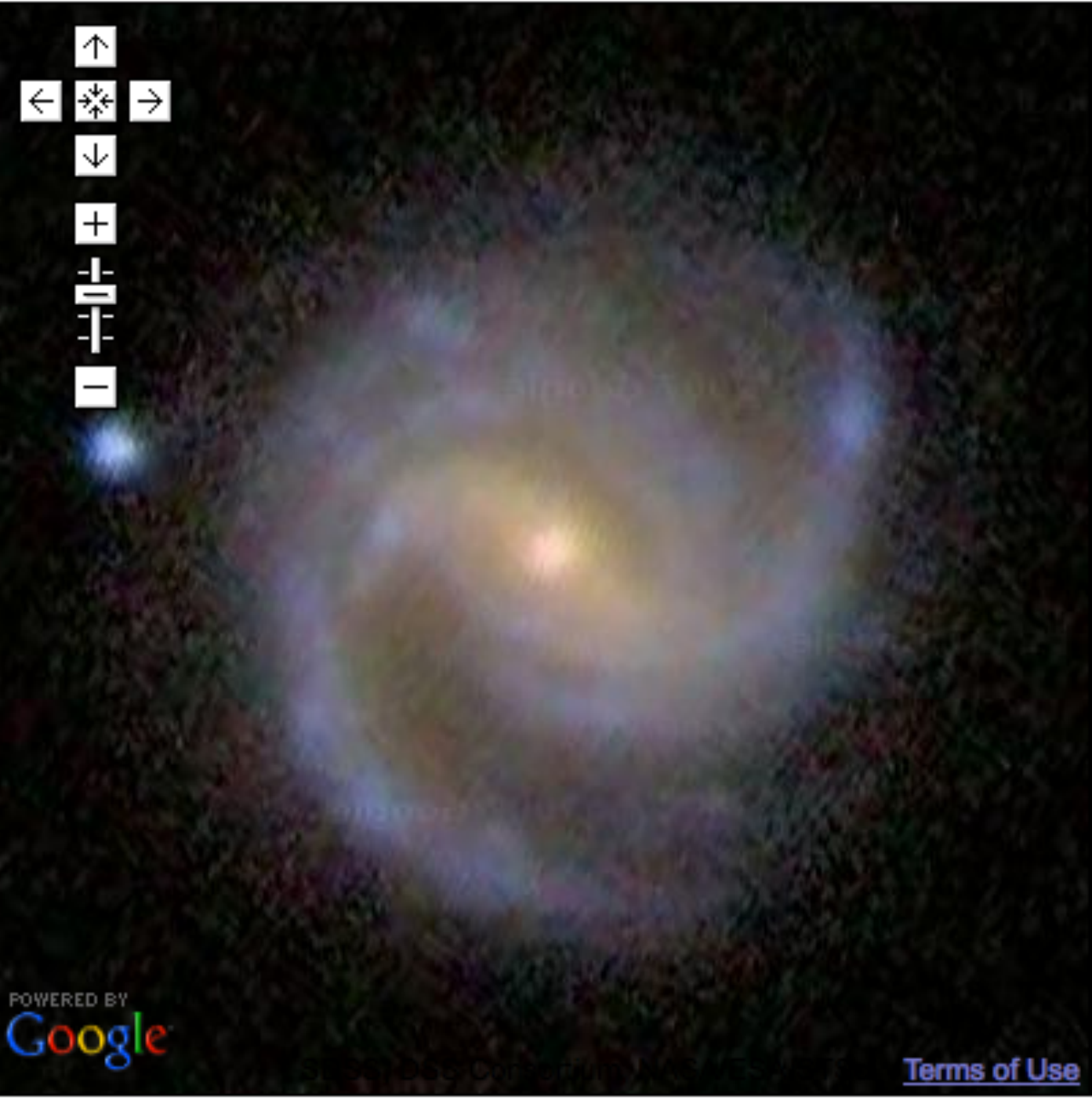}
   \includegraphics[scale=0.285]{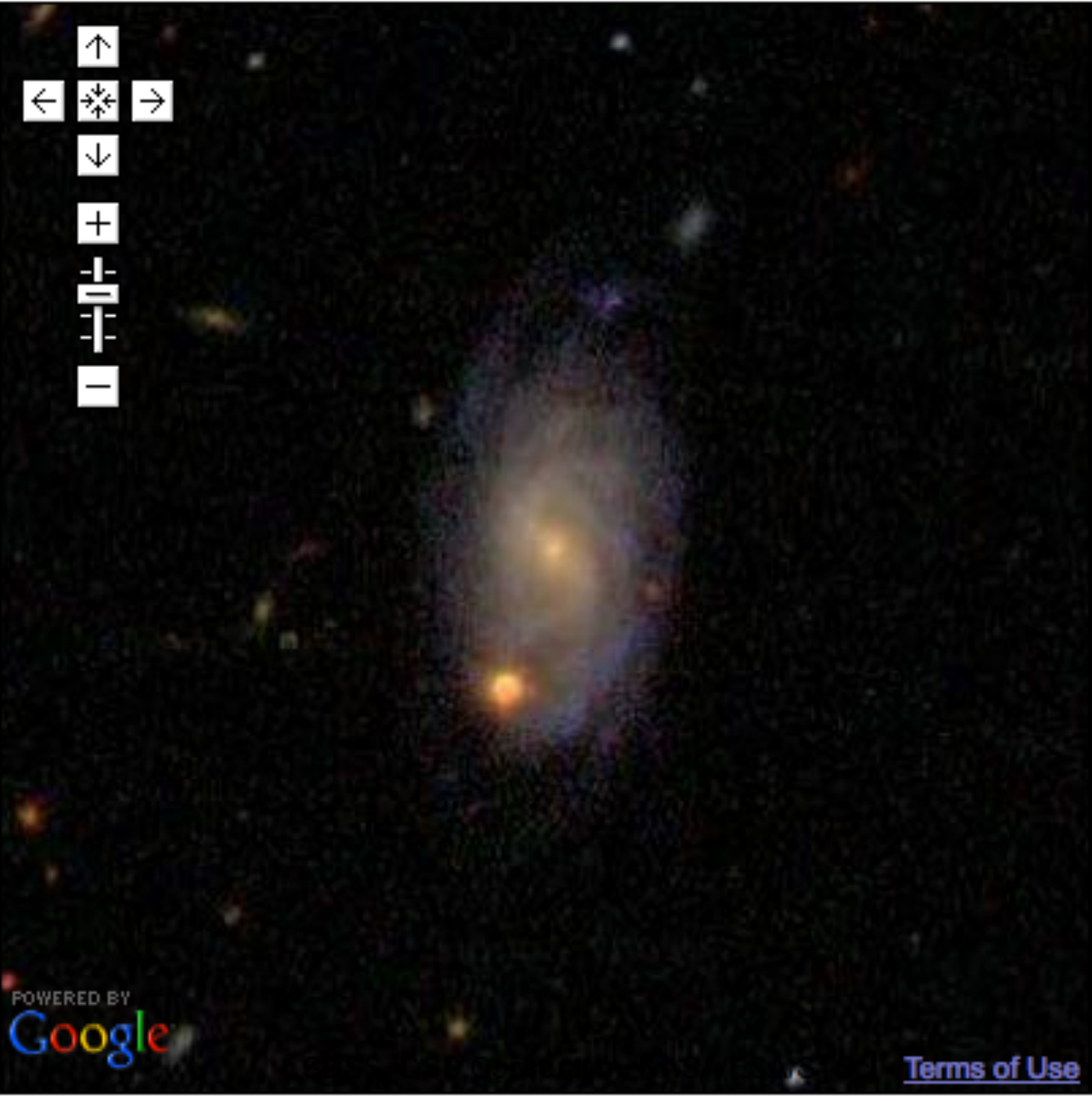}
      \includegraphics[scale=0.285]{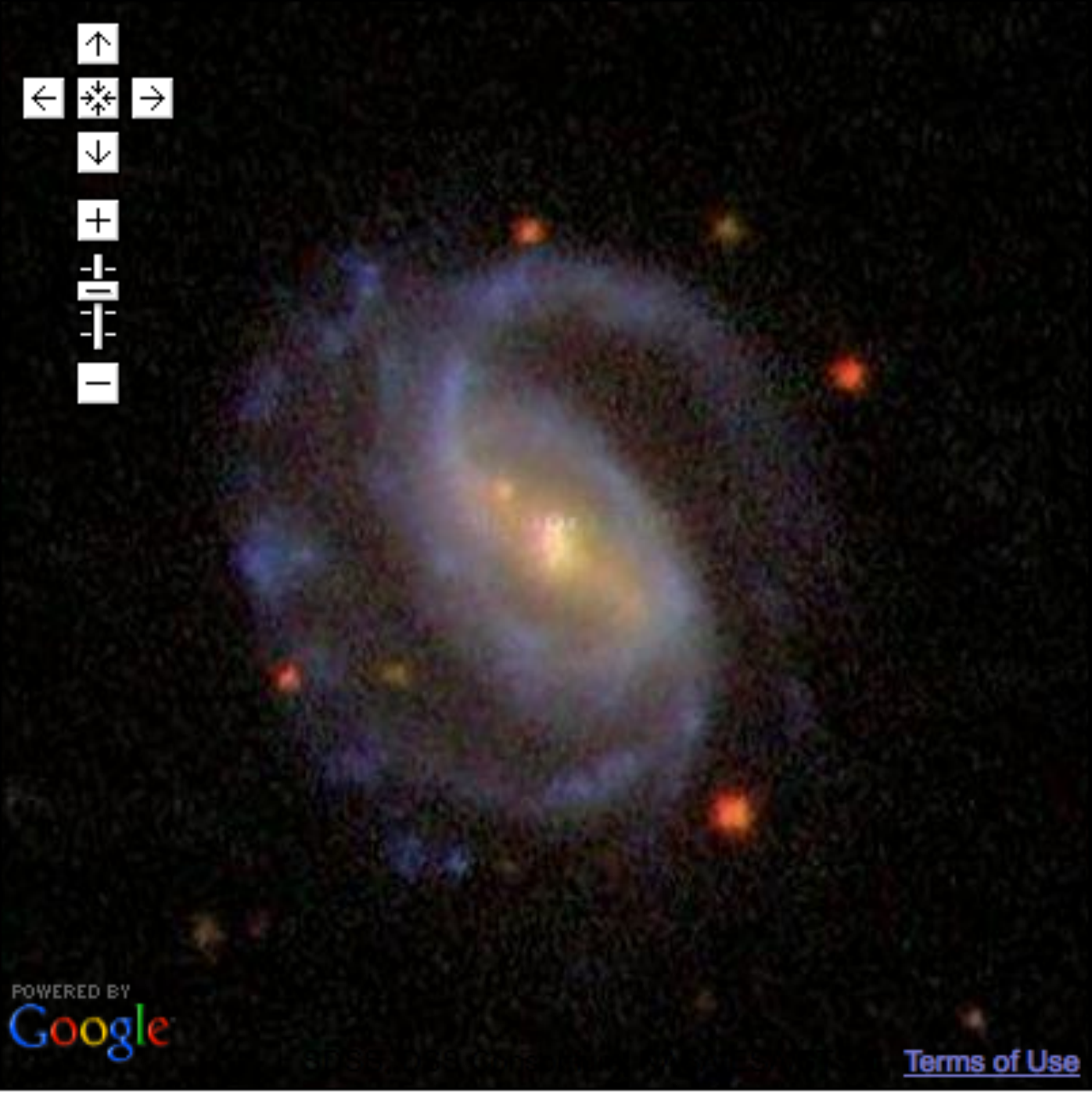}
         \includegraphics[scale=0.285]{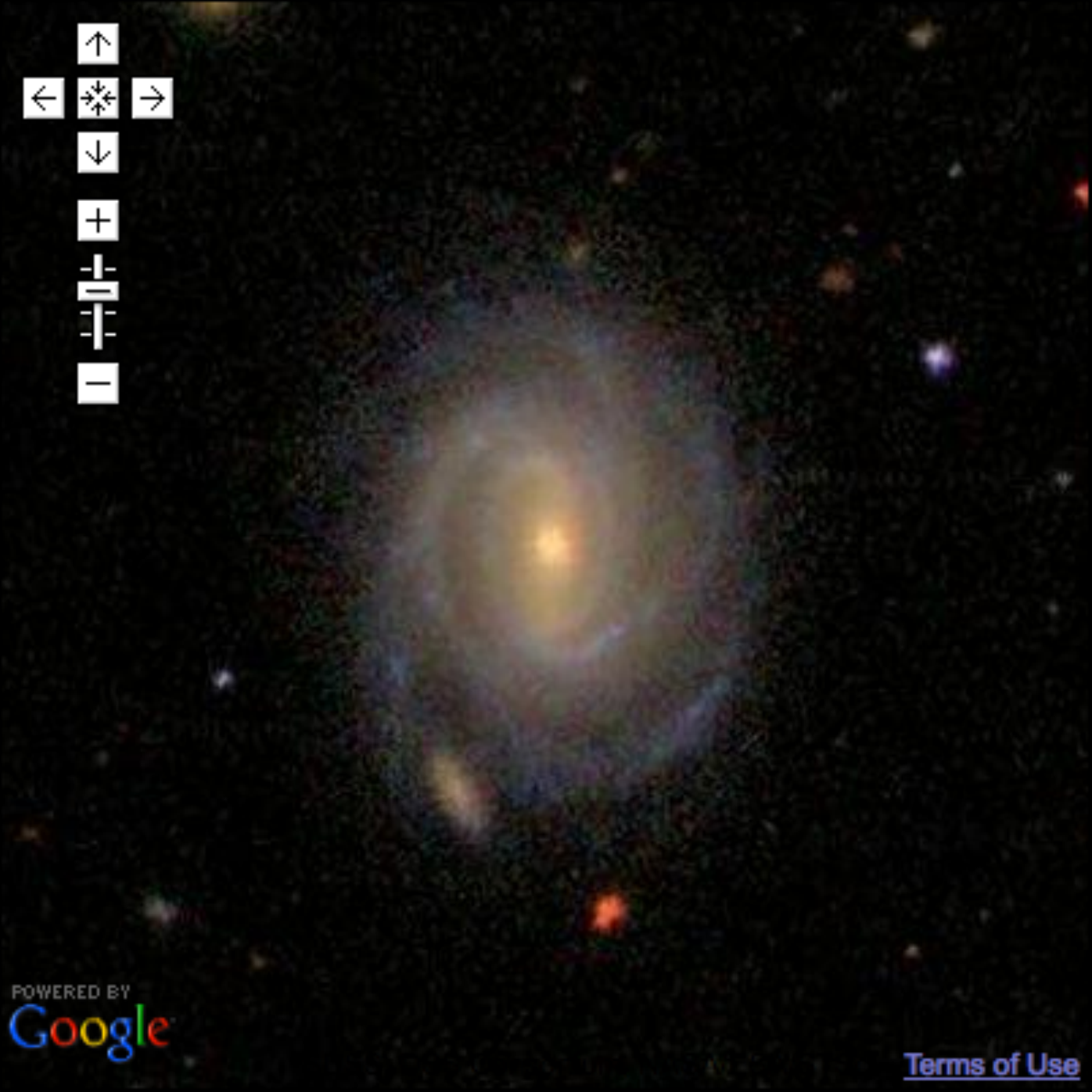} \\
   \caption{   \label{sprial_bar_example} Examples of the different  spiral arms  - bar connection with acronyms from Table \ref{bar_spirl_table} for GZ2 galaxies presented in the website's Google maps interface. (Row 1) No spiral arms or ring present, acronym NSR; (Row 2)  No spiral arms present, ring present, OR; (Row 3) Spiral arms connected to the ring, R; (Row 4) spiral arms connected to the bar, S; (Row 5) A mixture of bar and ring connection, SR. The galaxies were chosen to have $R_{Petro90}>15''$ for viewing considerations.}
\end{figure*}

We asked the observers to first identify if the SDSS photometric isophote\footnote{http://www.sdss.org/dr7/algorithms/classify.html\#photo\_iso}, shown as an ellipse plotted over the galaxy, was a good match to the galaxy shape. If not, the observer was asked to adjust the ellipse to better suit the galaxy image. The observer was then asked if a bar is apparent within the galaxy. If so, the observer marks the vertices of the bar, following a brief tutorial.  An ellipse was then drawn over the bar of length defined in the previous step, and the observer was asked to adjust the width of the ellipse to best fit the thickness of the bar. Finally, we asked the observer to indicate how the spiral arms were attached to the bar using one of the criteria shown in Table \ref{bar_spirl_table}. We show examples of the connections in Fig \ref{sprial_bar_example}. There are examples corresponding to the above questions and answers on a tutorial page connected to the website.

\begin{figure}
   \centering
   \includegraphics[scale=0.3]{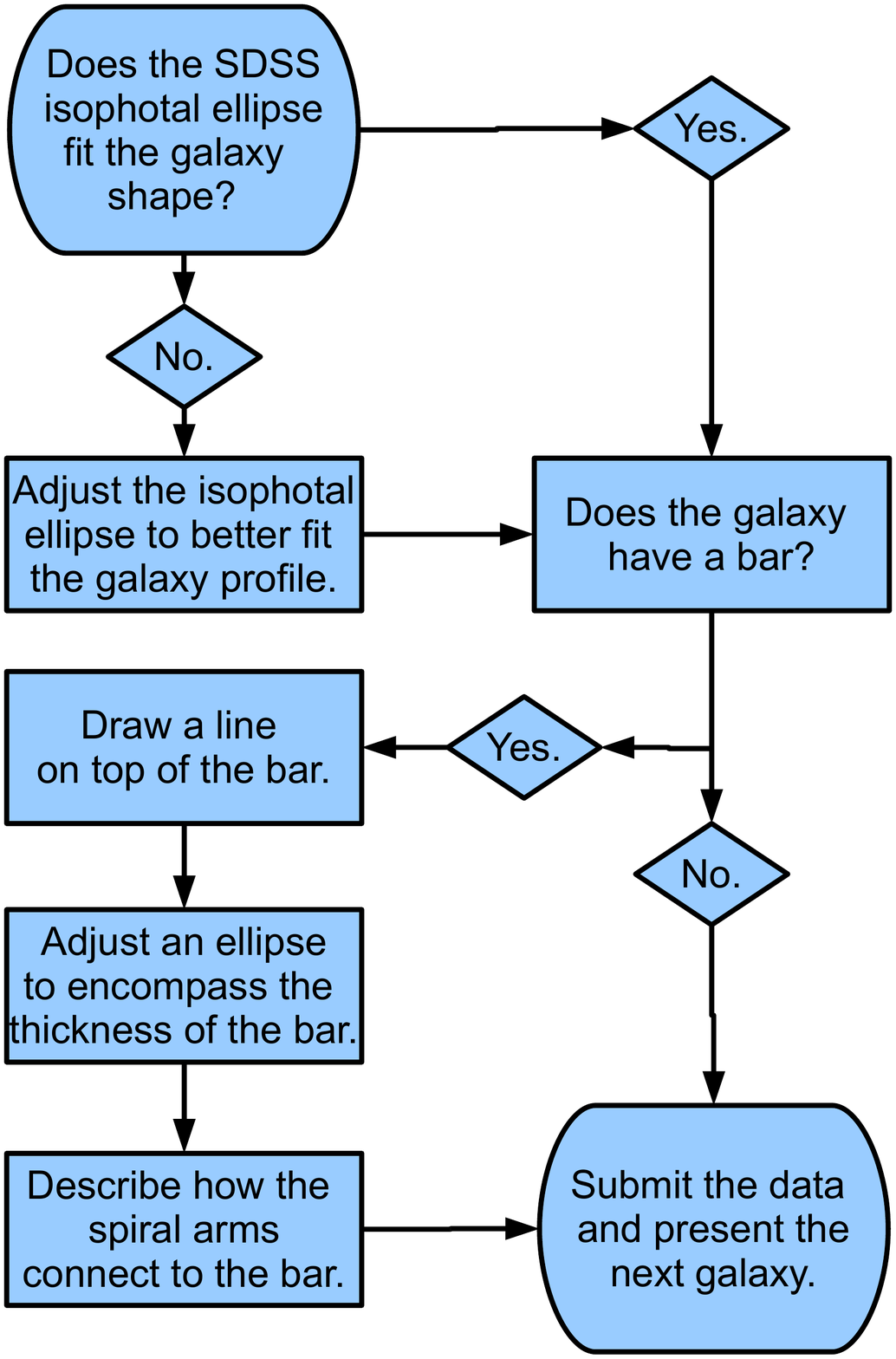} 
   \caption{ \label{flowchart}Flow chart of bar drawing website to identify barred galaxies and describe the bar properties.}
\end{figure}

\begin{figure}
   \centering
     \includegraphics[scale=0.4]{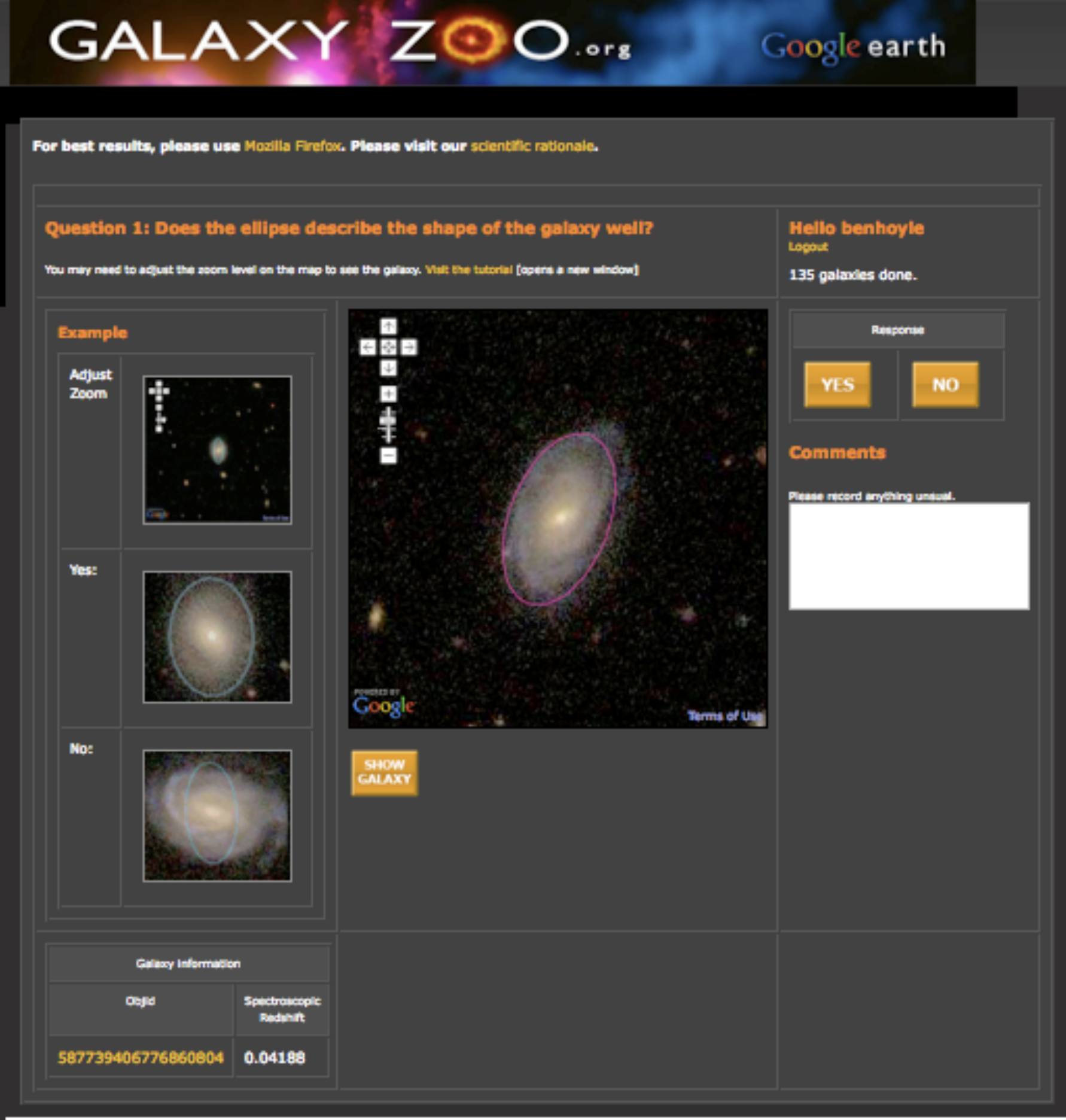}
   \caption{  \label{SiteScreenShot} A screen shot of the bar drawing website}
\end{figure}

A comment form allowed any interesting observations to be retained. A flow chart of the galaxy classification scheme is shown in Fig. \ref{flowchart} and a screen shot of the bar drawing website in Fig. \ref{SiteScreenShot}.

Once a galaxy had been classified a counter associated to it was incremented. The next galaxy to be classified was randomly chosen from galaxies with the lowest number of previous classifications.

\begin{figure*}
   \centering
     \includegraphics[scale=0.4]{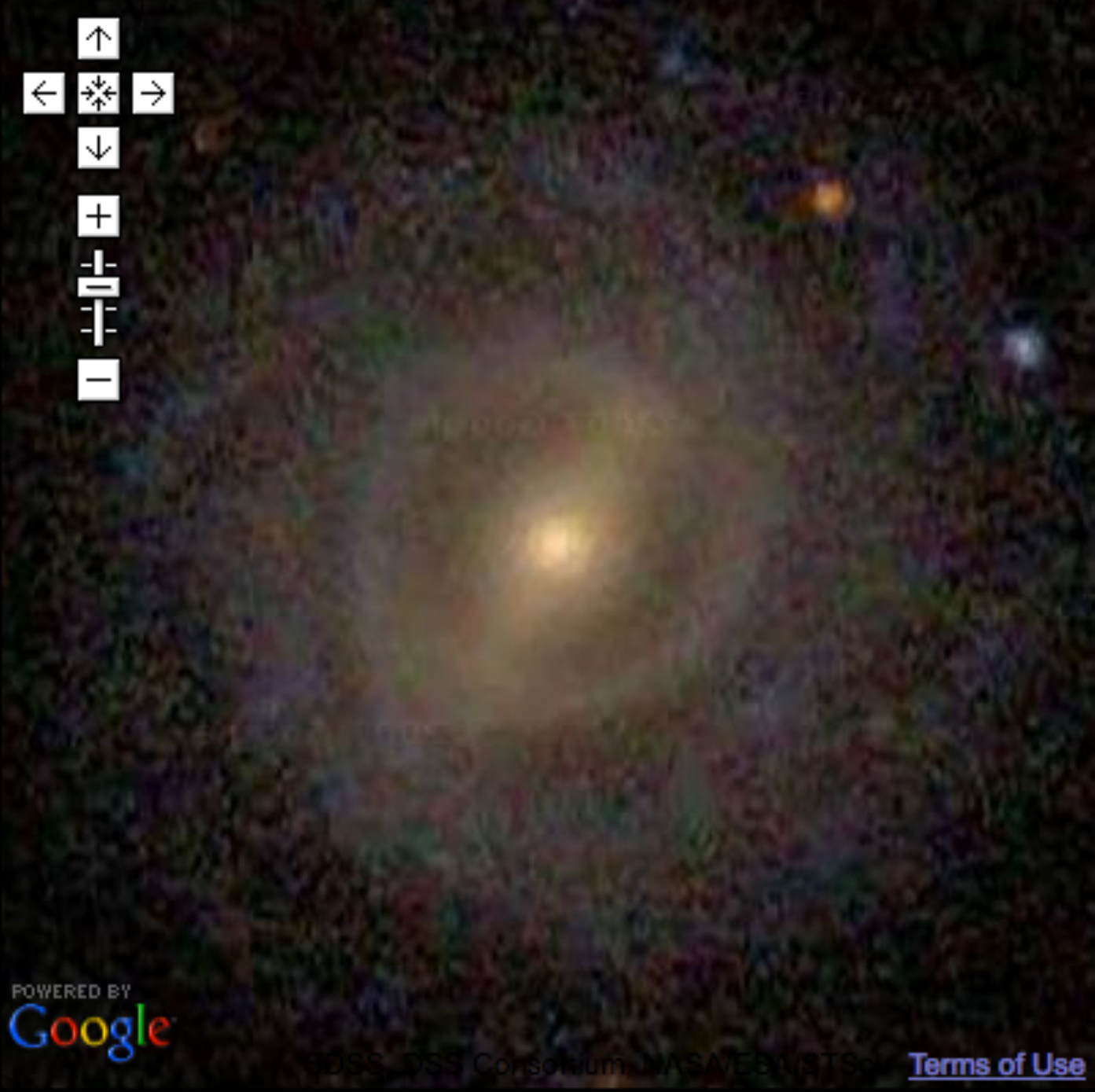} 
   \includegraphics[scale=0.4]{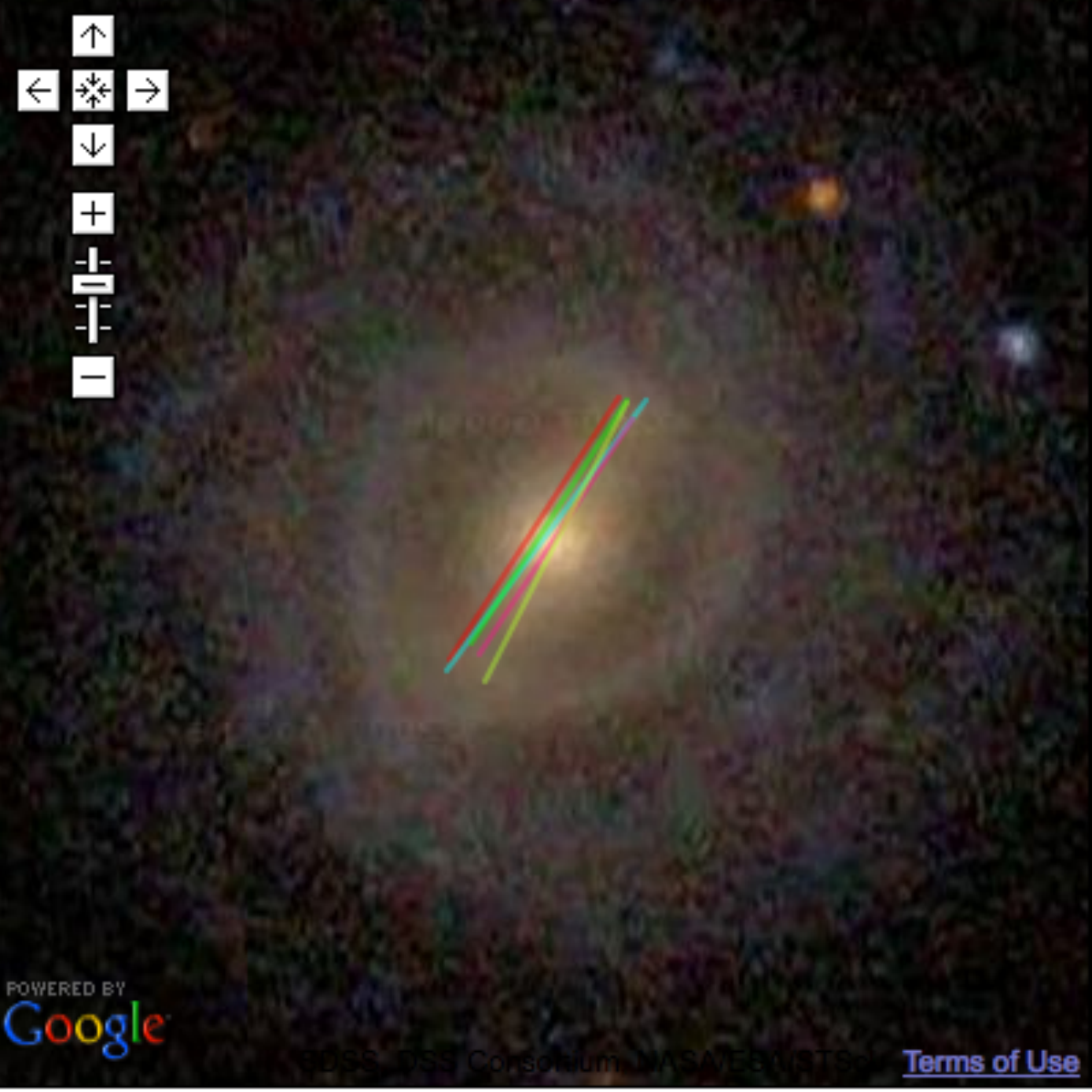} 
   \caption{  \label{barexample}An example barred galaxy as viewed in the Google Maps interface. The left (right) image shows the galaxy before (after) the independent bar drawings by $5$ observers. }
\end{figure*}

In Fig \ref{barexample}, we show an example galaxy in the Google Maps interface, before and after the bars have been drawn. This galaxy has been identified as having a bar each of the five times it was classified, and each bar has been drawn independently, without knowledge of the bar drawings of other observers.

\subsection{User Statistics}
\begin{figure}
   \centering
     \includegraphics[scale=0.25]{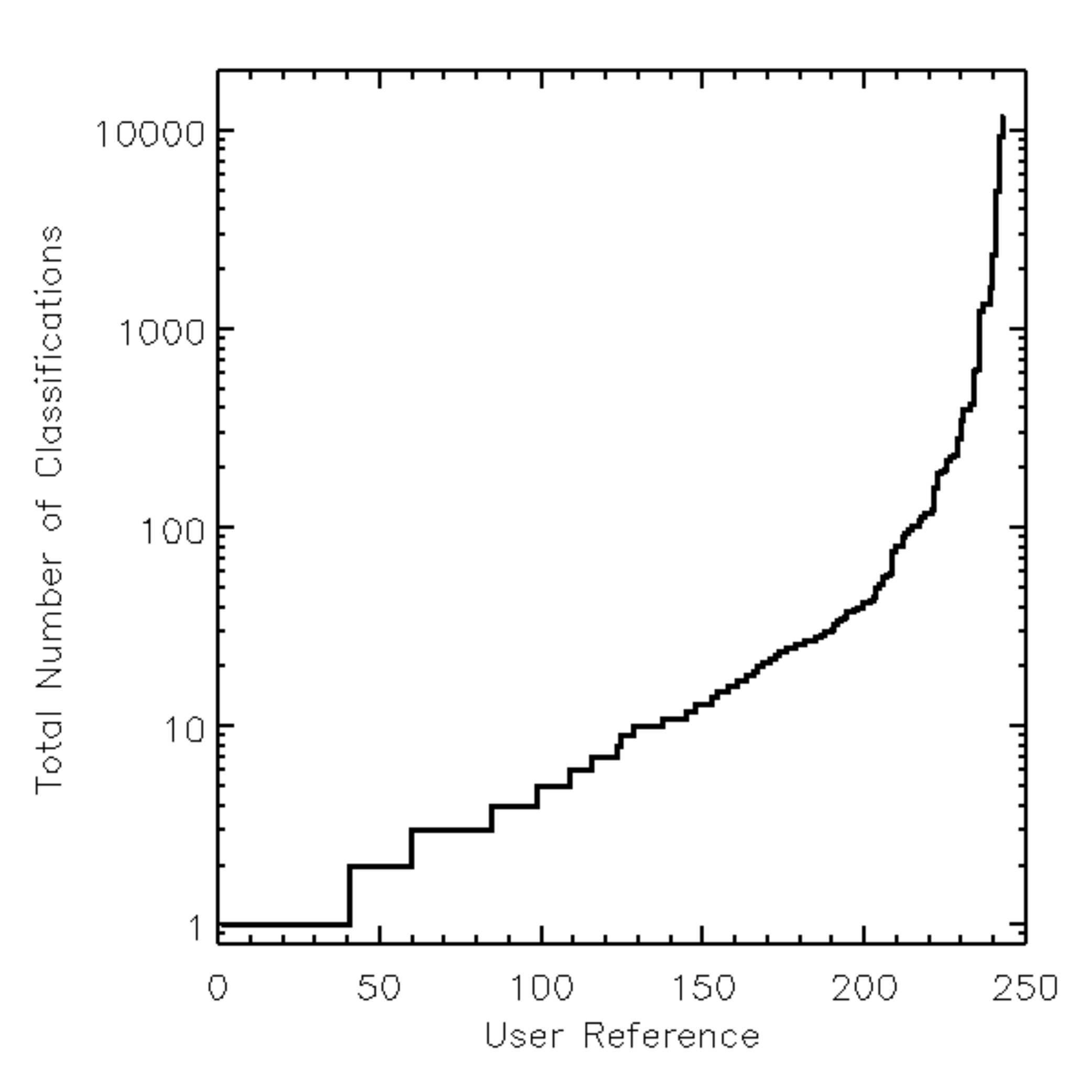} 
   \includegraphics[scale=0.25]{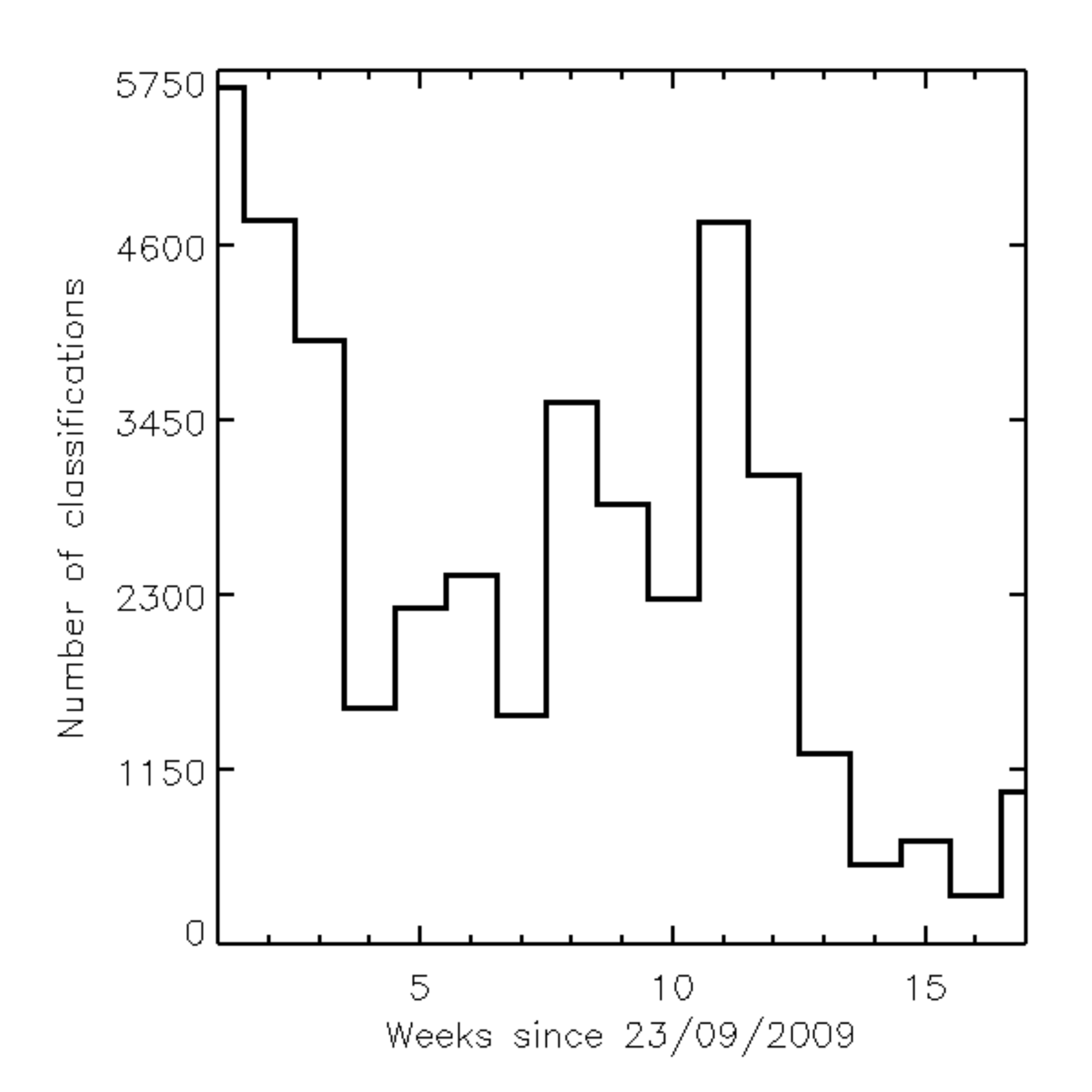} 
   \caption{  \label{userstats1}Histograms showing the number of classifications per individual observer (marked by an arbitrary ``User Reference"), ranked by number of classifications (upper panel), and the number of classifications performed each week since the launch of the site (lower panel).}
\end{figure}

In Fig. \ref{userstats1}, we show the number of classifications per week during the period we collected data, and the total number of classifications per person, ranked by classifications. The peaks in the time series, correspond partially to advertisements on the Galaxy Zoo blog\footnote{http://blogs.zooniverse.org/galaxyzoo/} and forum\footnote{http://www.galaxyzooforum.org/}. 

The Google Maps powered website was in operation from 23/09/2009 to 26/01/2010, and recorded a total of $45167$ unique classifications, of which $16551$ corresponded to bar drawings. Each galaxy was classified a minimum of five times, with $24.5\%$ were classified six or more times. 

We note that $50.4\%$ of the votes were cast by two observers (\emph{Graham Dungworth  \& Elizabeth Siegel}) and $78.0\%$ of the classifications were performed by seven observers (\emph{ oswego9050, Gravitroid, Elisabeth Baeten, Graham Dungworth, Caro, Elizabeth Siegel,Lily Lau WW}). Without the dedication of these observers, and the $242$ other observers (named in \S\ref{ack}), this project would not have been possible. 

\section{Data}
\label{data}
All the galaxies are drawn from the $\pi$ radians of the northern sky imaged by the Sloan Digital Sky Survey \citep[see][ and references therein]{York:2000gk,Gunn:2006tw,Smith:2002pca,SDSSDR7}, who derived photometric properties of $10^8$ galaxies, created color composite images, and took $10^{6}$ galaxy spectra. Subsequently, GZ2 asked observers, who had visited a classification tutorial, for detailed classifications of $250,000$ SDSS galaxies, including the presence (or not) of a galactic bar. For a complete overview of the GZ2 classification scheme see \cite{Masters:2010rw}.

\subsection{Input Galaxies}
The bar and control samples used in this paper were drawn from the July 2009 interim results of GZ2 classifications, which contained galaxies with more than ten GZ2 classifications at that time \cite[see][for a discussion of possible biases this cut introduces, which appear negligible]{Masters:2010rw}. GZ2 galaxies were selected from SDSS using the following cuts on Petrosian magnitude ($\le 17$), Petrosian radius ($\ge 3''$) and redshift ($0.005<z<0.25$). Due to image resolution restrictions, and the limited SDSS area available in Google Maps (mentioned in \S\ref{website}), we apply a redshift  cut of $z<0.06$ to all galaxies  
in the following samples, and remove galaxies in the SDSS coadded southern stripes.

We then select $5373$ galaxies which had been marked as containing a bar by at least $80\%$ of GZ2 observers who looked at them, hereafter the Control Bar Sample galaxies.  We include a control sample of $1000$ randomly selected disk galaxies which have been marked as containing a bar by $\le 5\%$ of observers, (the Control Non-Bar Sample galaxies), and $1000$ randomly selected galaxies from the full GZ2 sample with no constraint on bar probability (Random Sample I galaxies). We additionally included $1000$ randomly selected ``edge-on" galaxies, identified by greater than $99\%$ of GZ2 observers as being an edge on galaxy (Random Sample II galaxies). The total number of galaxies in the input sample, after chance duplicates were removed, was  $8180$.  We define ``bar length''  as the longest distance along the bar, i.e., from drawn vertex to drawn vertex, and was then used to fix the major axis of an ellipse. The ``bar width''  is defined as the minor axis of the ellipse, as adjusted by the GZ2 observer to best fit the bar. 

\subsection{Reliability of Measurements}
The random sampling of galaxies means some observers were presented with the same galaxy and therefore measured the same bar multiple times, allowing us to check both user consistency and inter user consistency. We define here the unitless  Bar length scatter, $\Delta L$, to be the bar length of all measurements per galaxy  $\vec{L}$, minus  the average bar length $<L>$, divided by the average bar length, e.g. $\Delta L = (\vec{L} - <L>)/<L>$.  This, and the bar length measurements, enable the following statistics to be constructed:

\begin{itemize}
\item{The bar length for each galaxy per unique observer, averaged over multiple measurements (if present). $L^u$}
\item{The average bar length for each galaxy, averaged over all observers and measurements, defined here as $L$ and used throughout the paper as simply ``Bar length" .}
\item{The standard deviation of the bar length scatter of each galaxy per unique observer, averaged over measurements.}
\item{The standard deviation of the bar length scatter \textbf{per galaxy} $g$, averaged over all length measurements of all observers,  $ \sigma (\Delta L_g)$.  }
\item{The average bar length scatter \textbf{per observer} $u$, averaged over galaxies they looked at and measurements they made, $\langle \Delta L^u \rangle$. }
\item{The standard deviation of the bar length scatter per observer, average over all length measurements and galaxies, $\sigma (\Delta L^u)$. }
\end{itemize}
Corresponding bar width statistics were also built.

\begin{table}
\begin{center}
  \begin{tabular}{c r r r } 
User ID & Ngals & $\langle \Delta L^u \rangle$ \% &  $\sigma (\Delta L^u)$  \% \\ \hline
$242$ & $691$ & $-3.5$ & $6.3$  \\ 
$243$ &  $580$ & $3.10$ & $6.2$  \\
$241$ & $392$ & $-9.7$ & $6.2$  \\
$240$ & $135$ & $4.7$ & $6.1$\\
$238$ &$32$ & $1.3$ & $5.5$ \\
$ 236$ & $25$ & $1.0$ & $6.2$ \\
$237$ &  $23$ & $-2.6$ & $5.3$  \\ 
$ 235$ & $6$ & $2.5$ & $5.6$ \\
$230$ & $4$ & $-2.1$ & $6.1$ \\ \hline
  \end{tabular}
\caption{\label{userstats} The intra-observer consistency of bar length measurements for the same galaxy. Columns show an observer identifier (as  the top panel of Fig. \ref{userstats1}), the number of galaxies that the observer has drawn bars upon more than once, the average bar length scatter per galaxy, averaged over all galaxies for each observer, and the standard deviation of the bar length  scatter per observer.}
\end{center}
\end{table}
In Table \ref{userstats}, we present statistics for those observers who multiple measured bar lengths on some galaxies. We show a observer identifier, the number of galaxies the observer had drawn upon more than once, the bar length scatter averaged over all such galaxies $\langle \Delta L^u \rangle$, and the average standard deviation of the bar length scatter averaged over all such galaxies, $\sigma (\Delta L^u)$. The averaged results of Table \ref{userstats} imply that observers are able to reproduce their own bar length measurements to $0.5 \pm 5.9 \%$ i.e. there is little bias and a scatter of $\sim 6\%$. 

\begin{figure}
   \centering
     \includegraphics[scale=0.3]{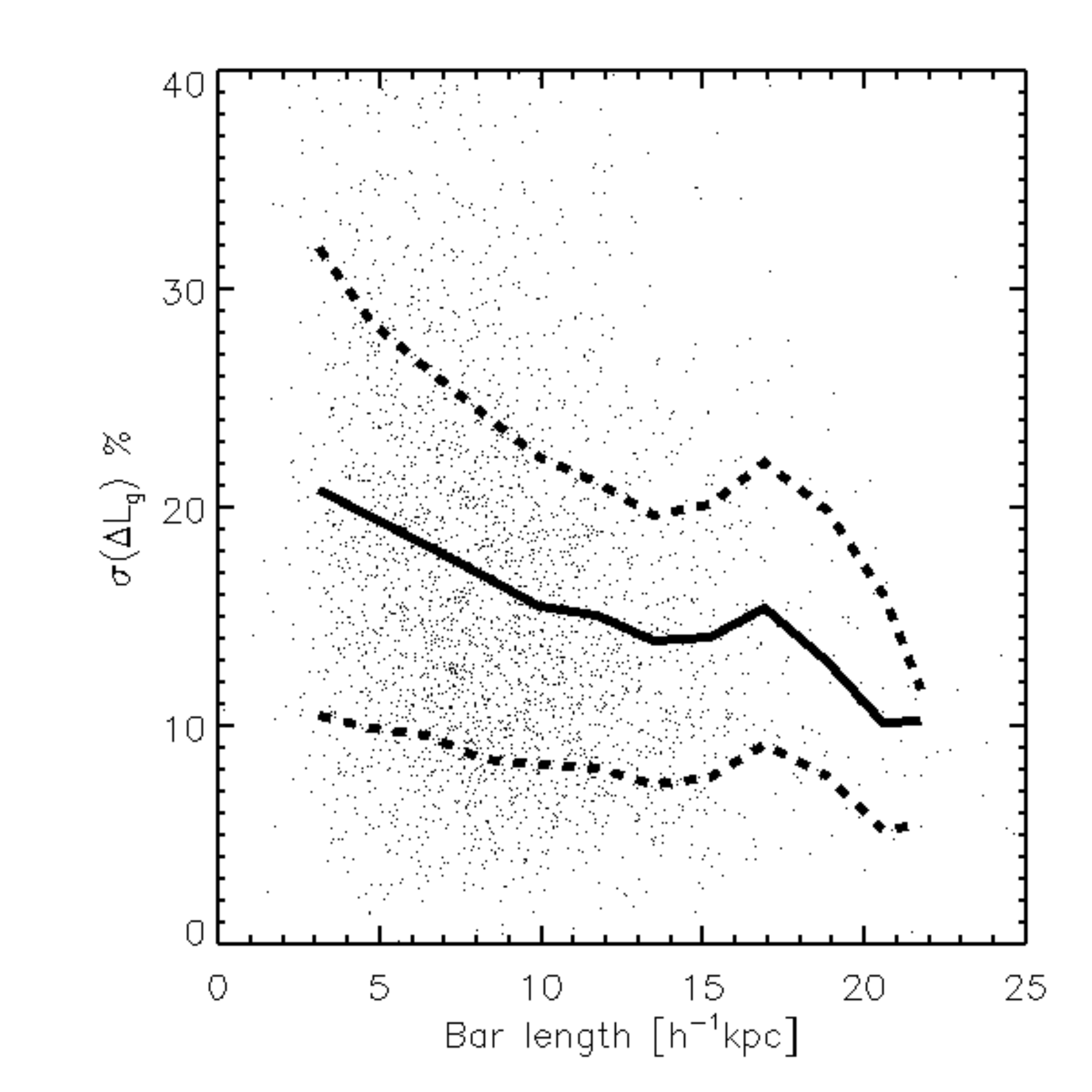} 
   \caption{  \label{baramstdev}The standard deviation of bar length scatter for each galaxy as a function of bar length. We plot the average value in the bin by the solid line and use the dashed lines to describe the $66\%$ spread of the data. }
\end{figure}

We next identify how well the observers can reproduce each others' (average) bar length measurements. In Fig. \ref{baramstdev}, we show the standard deviation of the bar length scatter per galaxy $\sigma (\Delta L_g)$ as a function of bar length, and see that observers are able to reproduce each others' average bar lengths, for each galaxy to $12\pm17\%$, and we find a slight improvement in agreement as a function of increasing bar length. For the two observers with the highest number of classifications, we find they reproduce each others' results to $10\pm 14\%$.

Of the $5373$ Control Bar Sample galaxies, $4911$ were identified as containing a bar by at least $1$ observer. Of the Control Non-Bar Sample galaxies, $284$ are marked as containing a bar. From the remaining $2000$ Random Samples I \& II galaxies, $169$ are stated by at least one observer as having a bar. We expect some of the randomly selected galaxies to contain a bar, because the random sample was drawn from all the GZ2 galaxies, independently of the number of bar identifications.  After visual inspection of a sample of the galaxies with bar measurements, we notice that occasionally bars have been drawn accidently, far away from the galaxy. To account for these spurious/accidental drawings and to build reliable statistics, we insist that at least $3$ bar measurements per galaxy have been made. The recovered Control Bar Sample galaxies now drops to $3195$, and the number of bar detections from the Random Samples I \& II galaxies drops to just $14$  (with only $1$ from the Random Sample II galaxies) and the number of Control Non-Bar Sample galaxies marked as containing a bar, i.e. ``false detections" drops to $22$. 

We have visually inspected the false detections, and they do appear to contain a weak bar in the Google maps and the SDSS Navigate interface. We have additionally examined the unrecovered barred galaxies, in the Google maps and SDSS Navigate interfaces, and the presence of a bar is indeed difficult to determine in the Google Maps Sky Interface and possibly dubious. We use the percentage of false detections ($2.2\%$) from the Control Non-Bar Sample galaxies to highlight the reliability of GZ2 observes to identify a barred galaxy relative to the GZ barred galaxy sample. In the following analysis, we only use galaxies with $\ge 3$ bar identifications.

\begin{figure}
   \centering
      \includegraphics[scale=0.3]{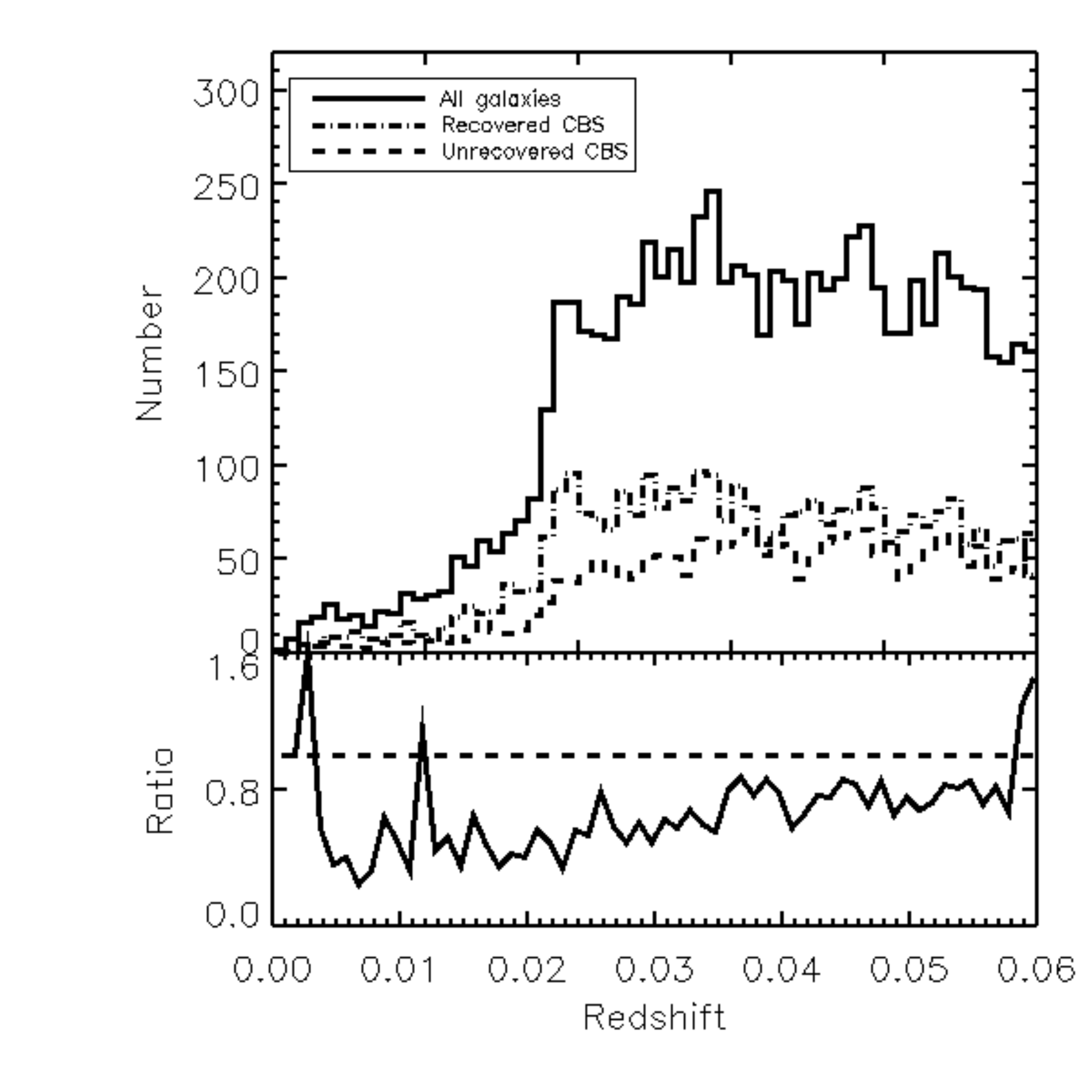} \\
    \includegraphics[scale=0.3]{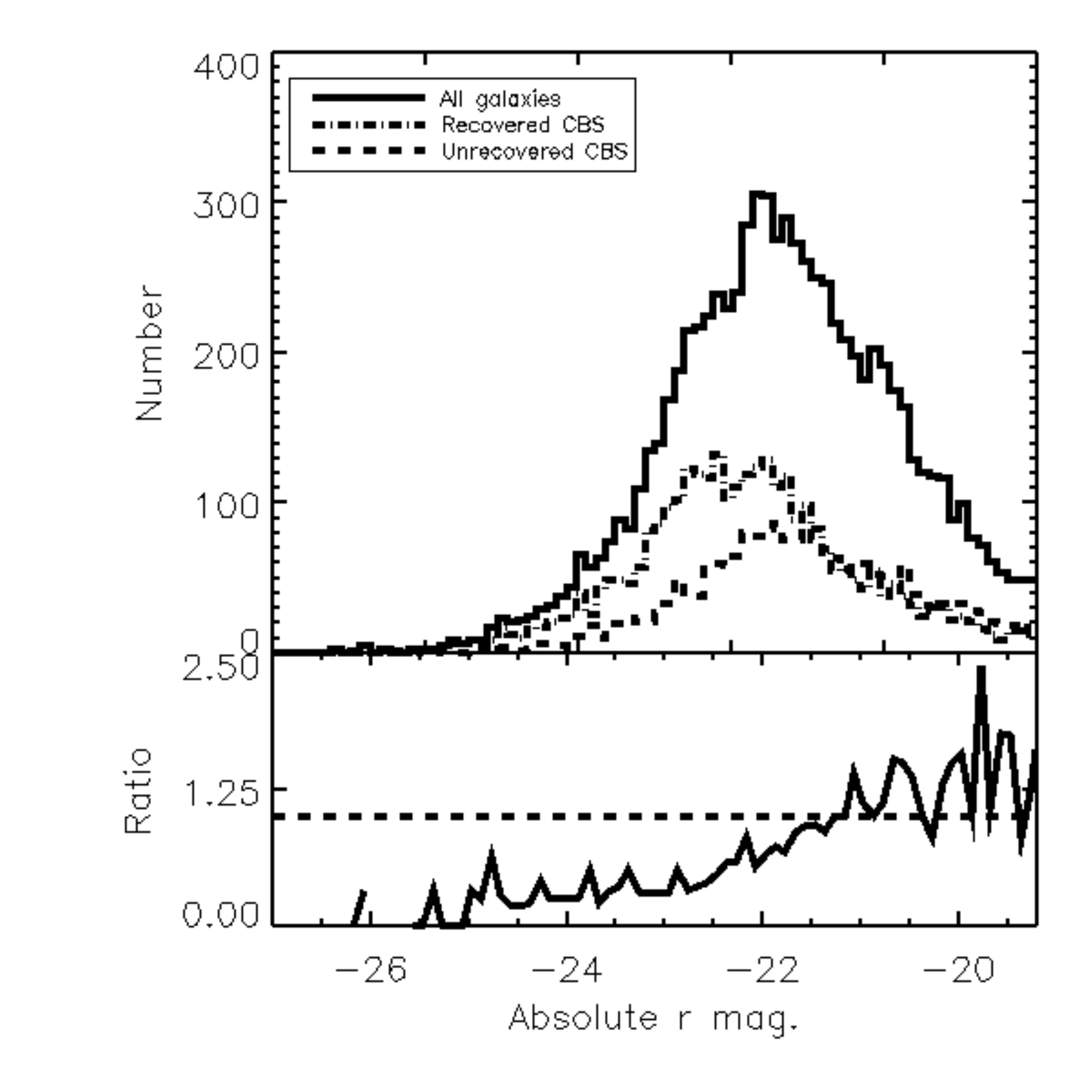} 
   \caption{  \label{Galpopsbars} We show the redshift (upper figures) and absolute $r$ band magnitude (lower figures) distributions of the populations of all $8180$ inputted galaxies (solid line), the Control Bar Sample (CBS) which were recovered by at least three observers (dot-dashed line) and the CBS which were not recovered by at least three observers (dashed line). The upper panels shows the distributions and the lower shows the ratio of the unrecovered to the recovered CBS, and the dashed line shows $y=1$.}
\end{figure}

In the panels of Fig. \ref{Galpopsbars}, we examine if there is a redshift or magnitude bias in the observers ability to correctly identify a barred galaxy from the Control Bar Sample galaxies.  In both upper panels, we show the redshift and absolute $r$ band magnitude distributions of the full $8180$ input sample, the recovered Control Bar Sample galaxies, and the unrecovered Control Bar Sample galaxies, and in the lower panels we show the ratio between the unrecovered and recovered Control Bar Sample galaxy samples. We see that at low redshift $z<0.02$ the ratio of the unrecovered to recovered  Control Bar Sample galaxies is  $0.4$, i.e., there are two times as many barred galaxies which have been correctly to be identified by at least $3$ observers, as there are incorrectly identified barred galaxies. The ratio increases to $0.8$ above $z>0.04$, which means the number of recovered barred galaxies is still greater than the number of unrecovered bar galaxies. This bias can be understood because lower redshift galaxies have larger angular sizes, and therefore it is easier to identify their substructure, such as bars.  Examining the ratio on the lower panel we see that the brightest galaxies ($M_r<-24$) which have bars are more easily identified than the fainter galaxies ($M_r>-22$). 

Finally, we cut on absolute magnitude ($M_r<-19.38$) to make a volume limited sample, and for plotting considerations, we remove four galaxies which have a ($g-r$) color $>2$. The final barred galaxy sample consists of $3150$ galaxies of which $99.1\%$ have been visually classified as spiral  galaxies and the remainder as early-type galaxies by the original Galaxy Zoo classifications \citep{Lintott:2008ne}.

\begin{figure}
   \centering
     \includegraphics[scale=0.3]{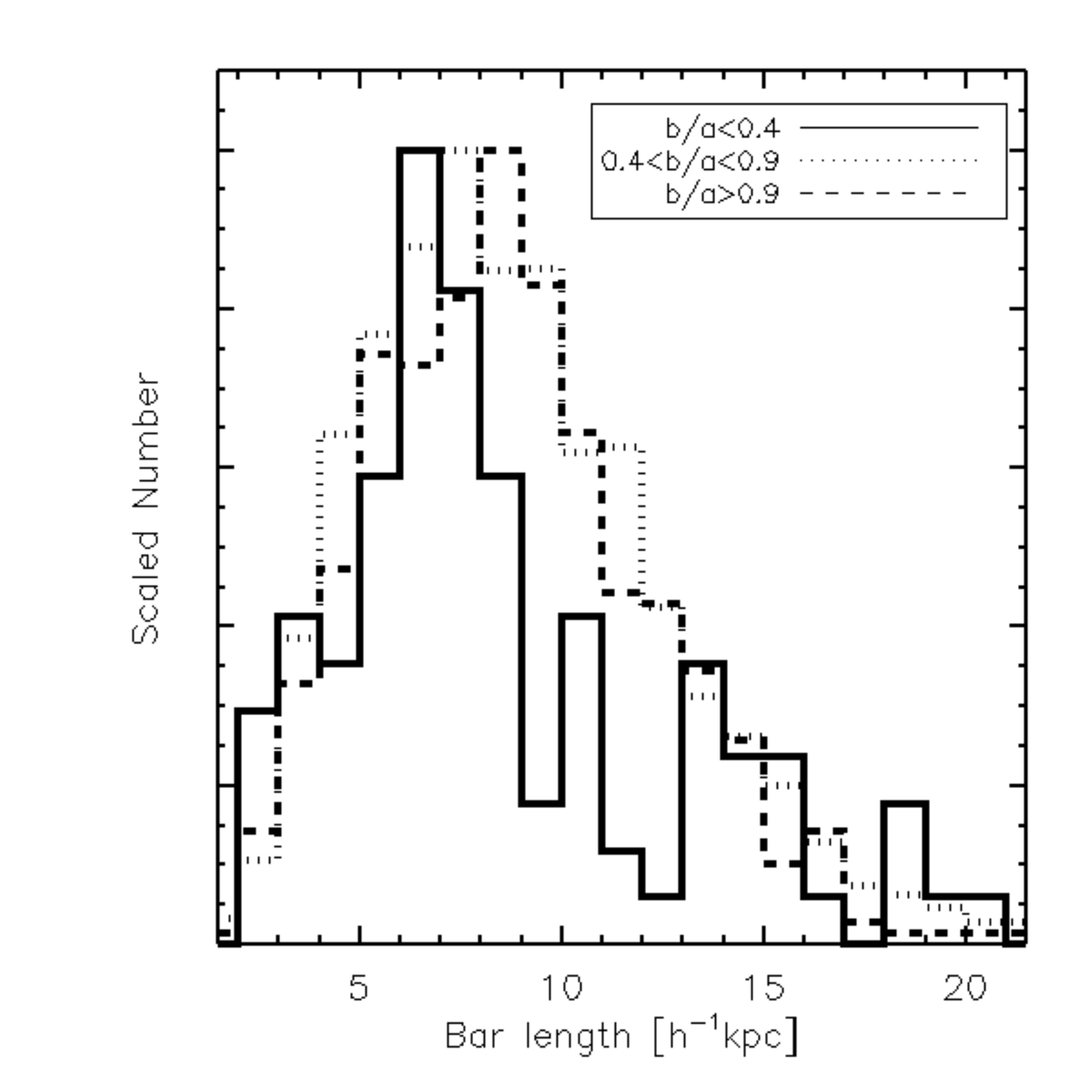} \\
   \caption{   \label{incl} The effect of galaxy axis ratio $b/a$ (as a proxy for galaxy inclination), on bar length. }
\end{figure}

We explored the effect of galaxy inclination on the measured bar length. As a proxy for the inclination we used the axial ratio of the SDSS measured isophotal ellipse axes, ($b/a$), and split the sample into three groups, galaxies which are highly inclined $b/a<0.4$,  slightly-inclined $0.4<b/a<0.9$, and  face-on $b/a>0.9$. In Fig. \ref{incl} we show the distributions scaled by number, as a function of redshift. We find that, as expected for randomly orientated galaxies, the majority ($2520$) of galaxies are slightly inclined.  We also find that the number of highly-inclined  (as approximated by axis ratio) galaxies is small ($102$) and the number of face-on galaxies is $528$. Comparing the scaled distributions across all bar lengths, we find that the slightly-inclined galaxies and face-on galaxies are very similar, but the highly-inclinded galaxies are associated with shorter bars perhaps because of projection effects.

\section{Correlating bar, disk, bulge and galaxy properties}
\label{bar_disk_prop}
In this section, we describe properties measured from knowledge of the bar lengths, and present correlations between bar and galaxy properties.

\subsection{Deriving Bar and Disk Properties}
The observers drew lines on the galaxy images measuring the bar lengths and we recorded the line vertices from each measurement. These were used to calculate the mean bar length and the mean distance between each bar vertex and the center of the galaxy. This allows an approximate measurement of the  $g-r$ ``bar" color which we define to be within the aperture from the center of the galaxy to the edge of the bar $0<R<R_{bar}$, and the ``disk" color, which we measure from the edge of the bar to the edge of the galaxy $R_{bar}<R<2\,R_{Petro90}$.  We define the edge of the galaxy (or ``galaxy size") to be two times the $r$ band Petrosian $90$ radius \citep[][]{Petrosian} as measured by the SDSS \citep{York:2000gk}. To measure colors, we determine the average $g$ and $r$ band flux of each galaxy, in each aperture, using the SDSS \emph{PhotoProfile \& ProfileDefs} tables and convert fluxes to magnitudes using $m=-2.5\,\log_{10}f$ \citep{1856MNRAS..17...12P}.  We note that colors measured in circular apertures are approximations of the bar and galaxy colors, but to first order we expect them to be representative. We define galaxy color as the difference between the SDSS $g$ and $r$ extinction corrected model magnitudes, and apply extinction corrections to the derived bar and disk colors \citep{1998ApJ...500..525S}. We also define a ``scaled bar length" to be the ratio of bar length to galaxy size.

Almost all $3150$ bars are significantly larger than the typical SDSS seeing of  $<2.5''$ \citep{Smith:2002pca}, only $28$ are close to this value ($<5''$). If the bars were of a comparable size to the seeing, they may become smeared and be undetectable. The smallest bars in our samples are $\sim 2 \,h^{-1}$kpc.

\subsection{Bar Length and Color}
\label{barcol_section}
\begin{figure}
  \centering
   \includegraphics[scale=0.3]{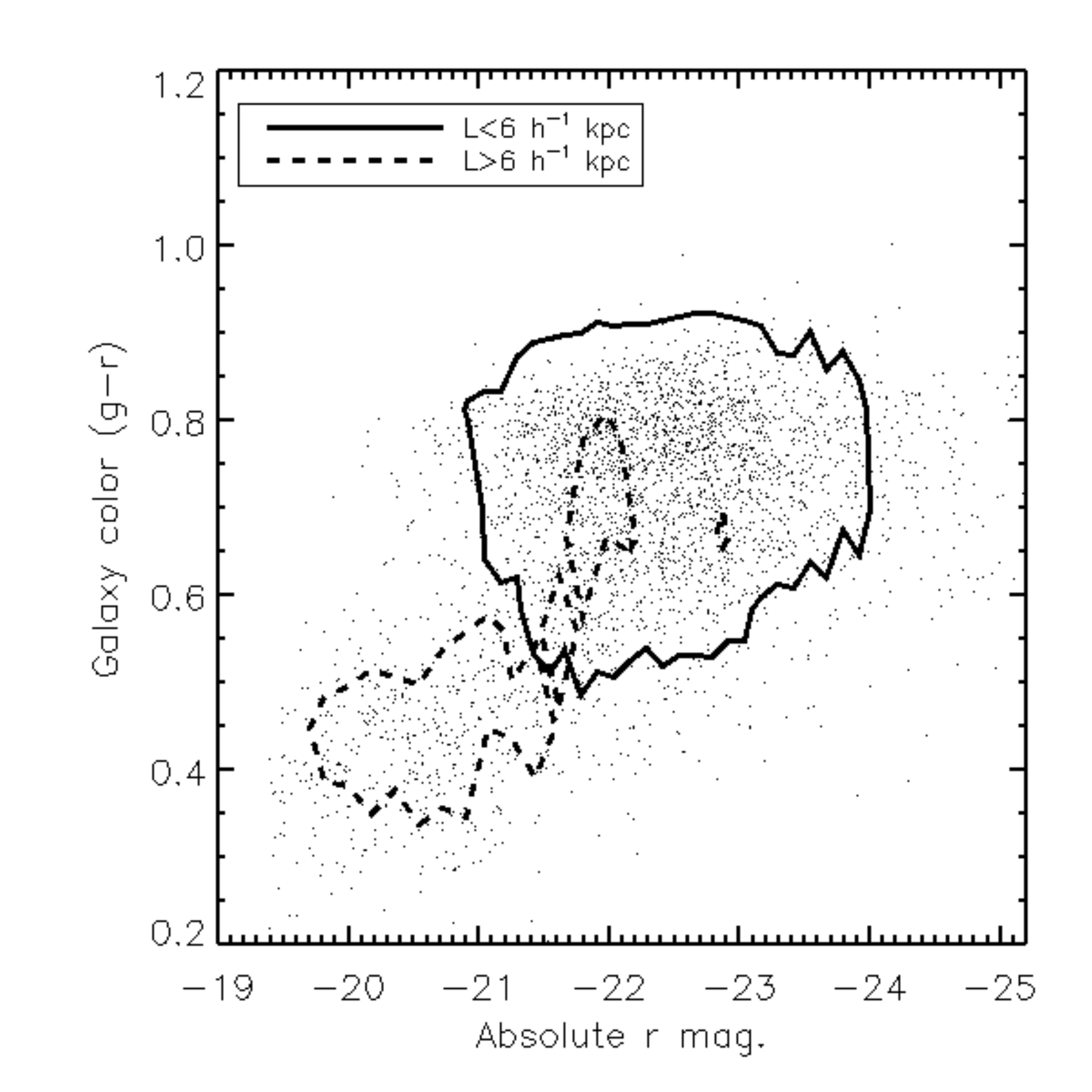} 
   \includegraphics[scale=0.3]{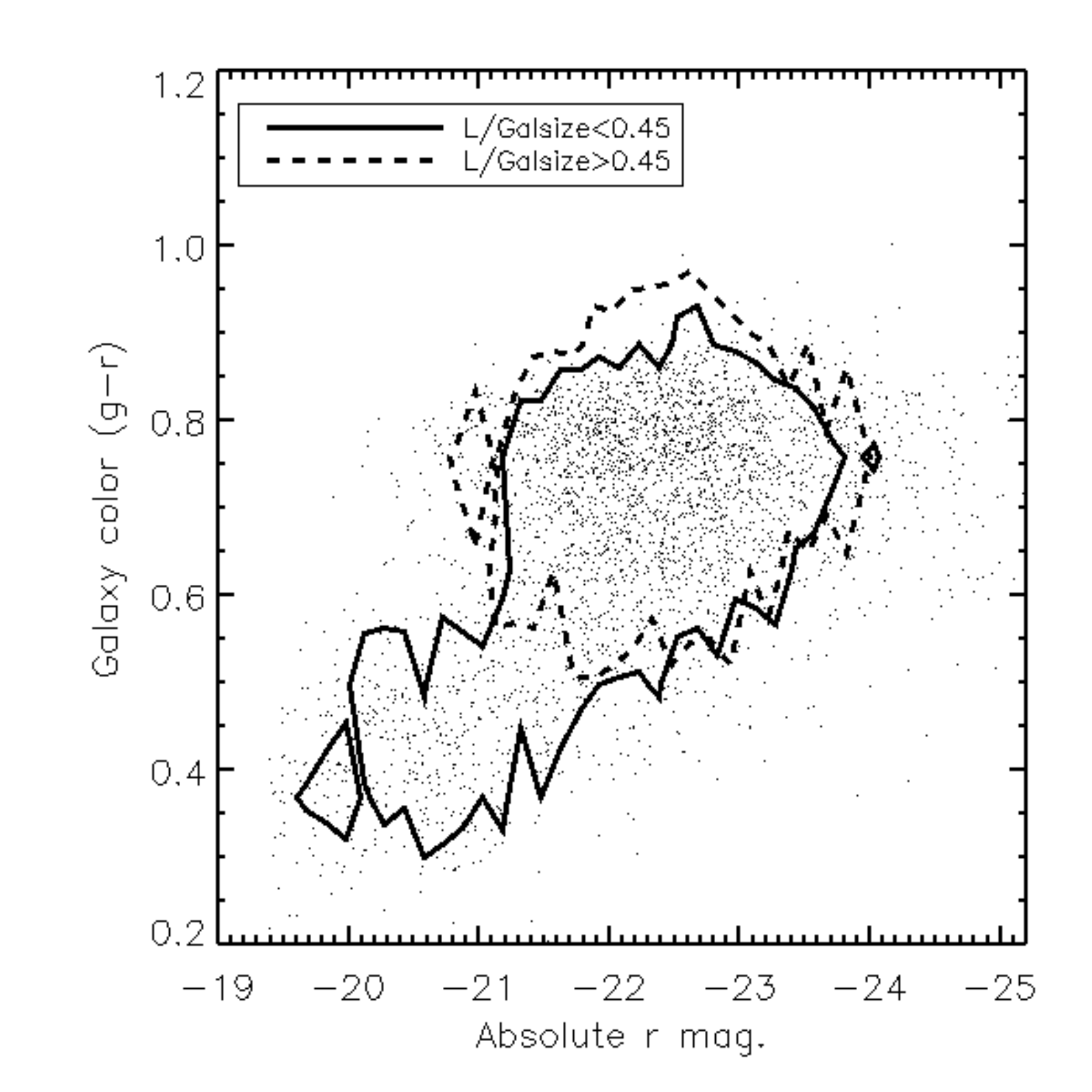} 
   \caption{   \label{colmagBarlength}The barred galaxy color-magnitude relation split by bar length. The top panel represents  isocontours of galaxies with bars lengths $>6\,\h^{-1}$kpc by the solid line, and bar lengths $<6 \,\h^{-1}$kpc by the dotted line. The bottom panel is the same as above, but shows isocontours of fractional bar length (bar length divided by galaxy size, see text) $<0.45$ by the solid line, and $>0.45$ by the dotted line.}
\end{figure}

In Fig. \ref{colmagBarlength}, we show the relationship between absolute $r$ band magnitude and galaxy color ($g-r$), and in the top panel, we over-plot isocontours of galaxy density with bar lengths greater than $6\,h^{-1}$kpc with the solid line, and less than $6\,h^{-1}$kpc using the dotted line. We find a clear segregation in the disk galaxy populations by color and magnitude (or bar length)  similar to the well known color-magnitude diagram for early and late-type galaxy populations \citep[see e.g.][]{Baldry:2003kj}. These disk galaxy subpopulations, split by color, have been seen before \citep[e.g., see][]{Cameron:2010he,Nair:2010xh,Masters:2010rw}, but this is the first time it has been shown as  function of bar length. The bottom panel is the same as the top, but shows isocontours of fractional bar length (bar length divided by two times the $r$ band Petrosian radius $90$ as a measure of galaxy size) $<0.45$ by the solid line, and $>0.45$ by the dotted line.


\begin{figure}
   \centering
     \includegraphics[scale=0.3]{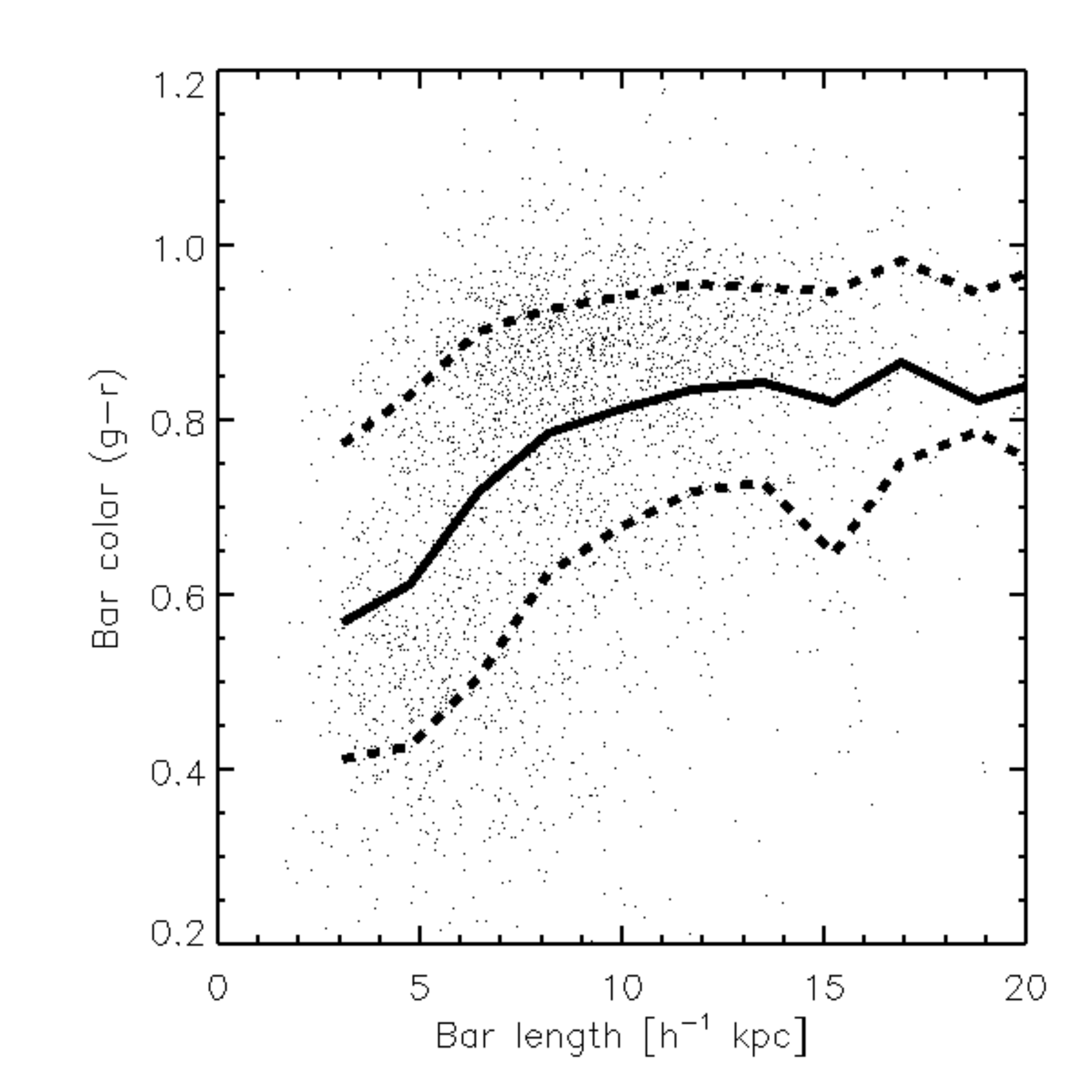} \\
   \caption{   \label{barcolbarlength} The effect of bar length on bar color. We find shorter bars are bluer than longer bars, and bar color increases with redness, until a critical length (or color), after which the change in color is reduced.}
\end{figure}

In Fig. \ref{barcolbarlength} we show the bar length against bar color (recall, this is defined as the average color of the galaxy interior to the bar radius) and plot the average value in the bin by the solid line and use the dashed lines to describe the $66\%$ spread of the data. We find that bars which are short in absolute length $L<6\, \h^{-1} \,\kpc$ (and also in scaled bar length, $L/2\,R_{Petro90} < 0.3$) are bluer than longer bars. As the bar length increases, the bar color becomes redder until $L \sim 10\, \h^{-1}\, \kpc$ (or $L/2\,R_{Petro90} \sim 0.5$) at which length the colors become constant.

One may worry that this correlation could be a systematic bias, in that longer bars are predominantly found in redder early-type disk galaxies because star formation has obscured some of the bar in bluer late-type disk galaxies.  However, this would be contrary to the work of \cite{Sheth:2007cp} who find that the barred fraction of a sample of low redshift galaxies is constant across the SDSS  $g,r,i,z$ bands. Additionally, \cite{Erwin:2005si} compiled a collection of $135$ barred galaxies with optical and near infrared imaging, which is less sensitive to obscuration by star formation, and found that the average bar length in early-type disk galaxies is $2.5$ times that of late-type disk galaxies. Therefore, we argue this correlation is physical, in agreement with \cite{Erwin:2005si} and not due to a systematic bias.

\subsection{Galaxy Color and Bar Color}
\begin{figure}
   \centering
 \includegraphics[scale=0.3]{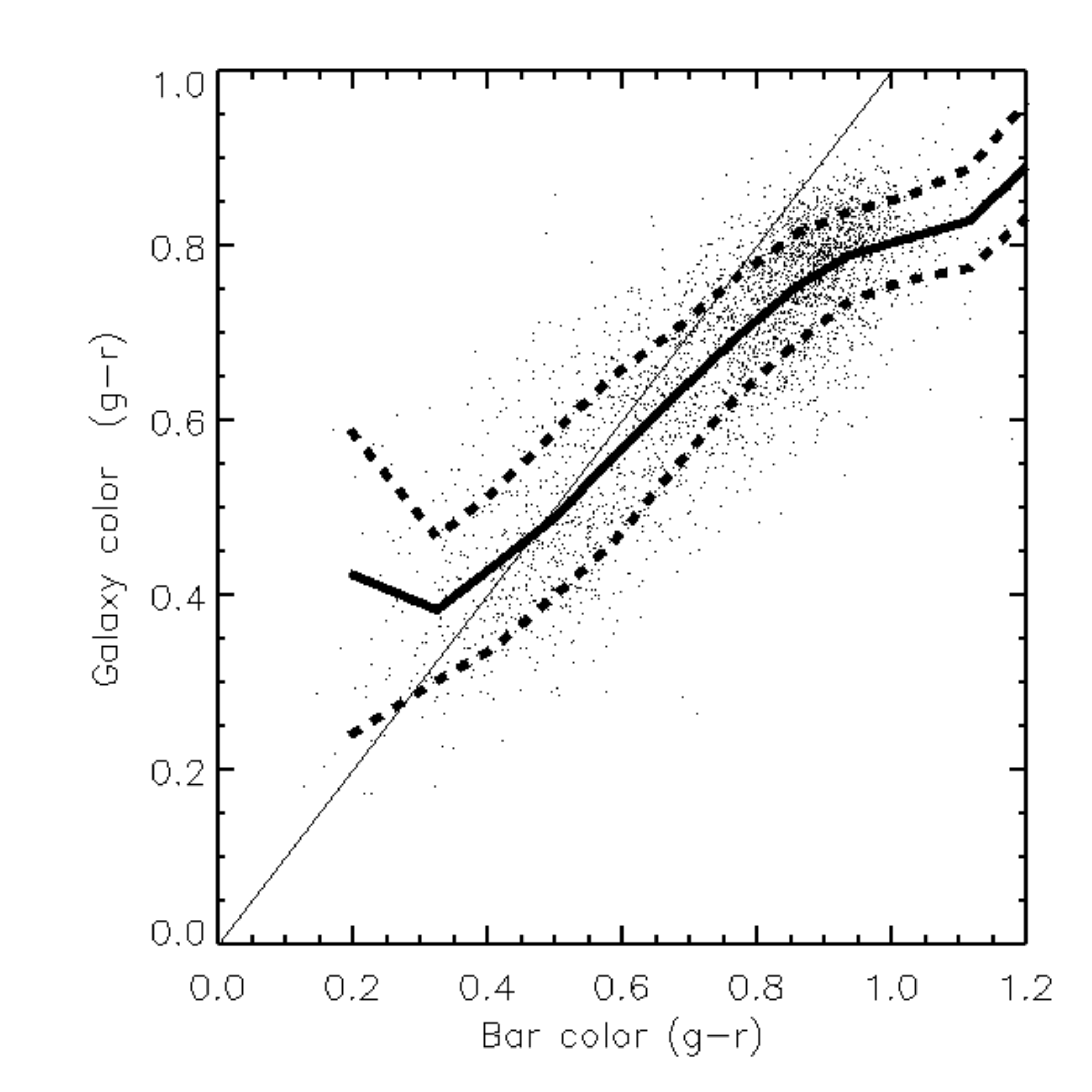}
  \includegraphics[scale=0.3]{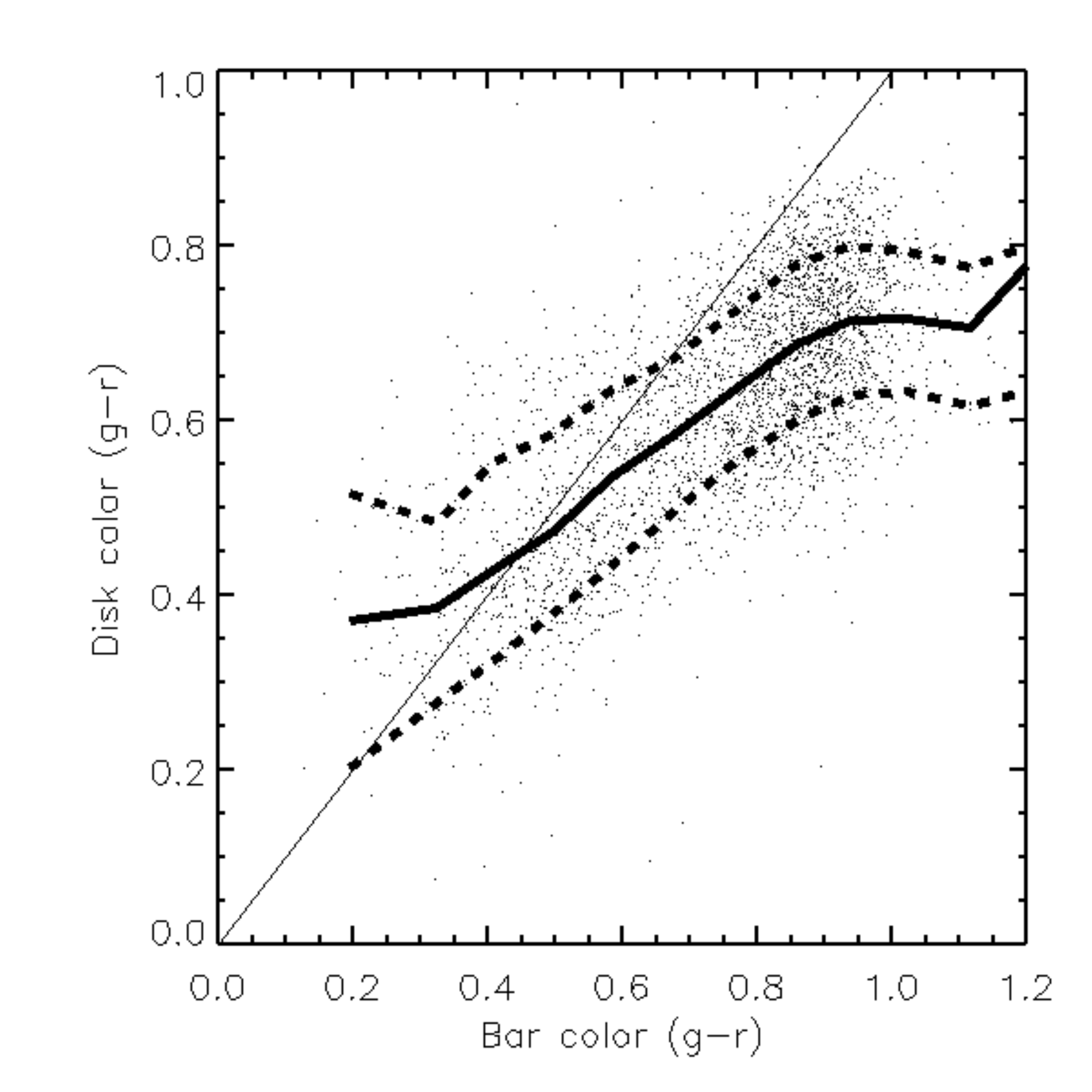} 
   \caption{ \label{bardiskcol} We show the bar color against galaxy color (disk color) upper (lower) figure, and also plot lines of equality. Both the total galaxy color and disk color become redder as the bar color becomes redder.}
\end{figure}

 In  Fig. \ref{bardiskcol} (upper panel) we show the color of the bar and the galaxy color (as measured by the SDSS), and find that, as one might expect, as the bar color becomes redder, the total galaxy color becomes redder. In Fig. \ref{bardiskcol} (lower panel) we examine the relationship between bar color and disk color.
 
We see that bar colors and disks colors are highly correlated, as expected from simulations \citep[][]{Scannapieco:2010nw}. Interestingly, the bluest (therefore smallest) bars have disks which are redder than the bar color, but for all other bar lengths, the bar colors are redder than disk colors. This can be understood if, as the bar increases in length, some process shuts off star formation in the entire galaxy from the inside out. The low scatter implies that the stellar populations of both bar and disk are drawn from similar parent populations. Further exploration of the relationship between the change in bar, disk and galaxy color as the bar length increases will be the topic of a future publication. 

\subsection{Bulge Prominence}
\begin{figure}
   \centering
   \includegraphics[scale=0.3]{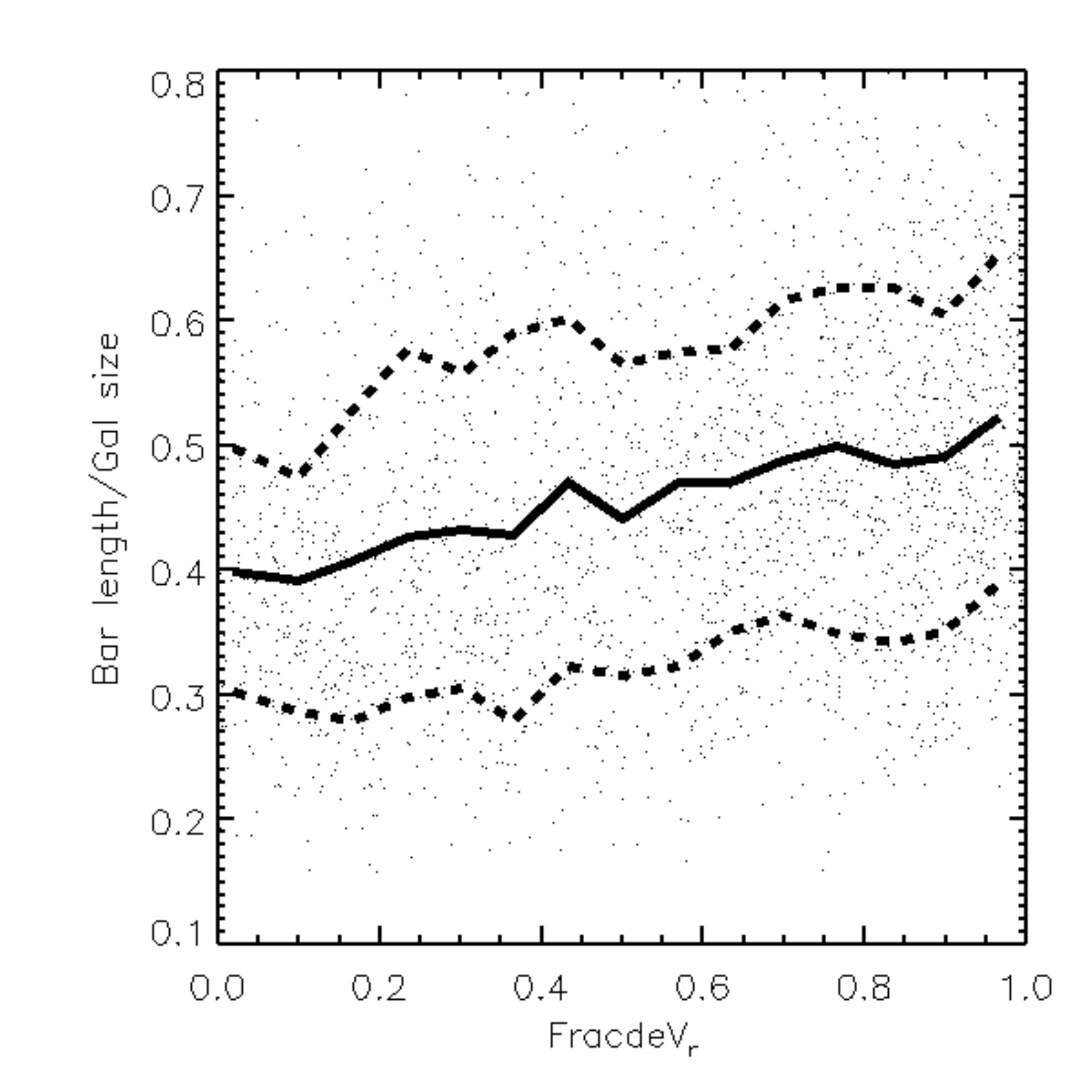} 
   \caption{   \label{fracdevbarlength}The {\tt fracdeV}  (or prominence of a the central galactic bulge.) plotted against bar length/galaxy size. As  {\tt fracdeV} increases the bars become longer.  }
\end{figure}

As suggested by \cite{Masters:2010rw}, we use the SDSS measured $r$ band {\tt fracdeV} as a proxy for bulge size and show {\tt fracdeV} as a function of bar length, divided by galaxy size, in Fig. \ref{fracdevbarlength}.

As the bulge increases in size, the scaled bar length (bar length divided by galaxy size) also increases, albeit with a large scatter. Fitting a line to these points we find a correlation with a slope $0.12\pm0.05$, i.e. the bar reaches $12\%$ further into the disk in galaxies with a large bulge compared to galaxies without a central bulge. This observational relationship is not new, and is expected from simulations \citep[e.g.][]{G1980A,Gadotti:2010ct,Athanassoula:2003wd}, but we test it here with a large galaxy sample.  We note that, in absolute bar length ($h^{-1}$kpc), galaxies with  {\tt fracdeV} $=1$ have on average, bars which are $2 \pm1.5$ times as long as galaxies with {\tt fracdeV} $=0$.  This may point to a systematic error as longer bars may influence the SDSS {\tt fracdeV} fits, however \cite{Masters:2010rw} showed little impact of bars on {\tt fracdeV} distributions.

\subsection{Spiral Arm Connection to Bar}
We next divide the galaxies into subsamples of the different possible connections between the spiral arms and ring (if they exist) and the bar, by using the results of the final question on the classification site.  In Fig. \ref{sprial_bar_example}  we showed a random selection of four galaxies per classification criteria as seen in the Google Maps interface with rows describing, from top to bottom, NSR (Spiral arms and ring are not present), OR (Spiral arms are not present
and the bar is connected to the ring), R (Spiral arms are connected to
the ring around the bar), S (Spiral arms are connected to the end of the bar), SR (Spiral arms are connected to a mixture of a ring and the bar). 

The connection measurement is difficult to perform, being potentially subjective, and is further confounded by the limited resolution of the galaxy images in the Google Maps interface (see \S\ref{website}). We tackle these issues by continuing along two parallel threads. In the first (our ``clean sample'') we only consider galaxies where different observer classifications agree to a high significance, this cuts the sample size from $3150$ to $771$. We later use the full sample and weight the number of classification per category per galaxy by the total number of classifications per galaxy. This allows each classification for each galaxy to be used statistically. In what follows, we compare bar lengths, widths (or strengths) with bar and galaxy properties for each of the connections outlined in Table \ref{bar_spirl_table}. 

\subsubsection{Clean sample}
\label{cleangals}
We collate votes for each of the six possible outcomes (as described in Table \ref{bar_spirl_table}) with the constraint that the inter-observer agreement must be $100\%$ for galaxies with three bar measurements, and $80\%$ for galaxies with greater than three bar measurements. This cleaning reduces the number in the sample to $771$, which are distributed per category as shown in Table \ref{conn_res_cl}. We also show the fractional dispersion for bar length measurements per galaxy, averaged over all galaxies, and the standard deviation of the fractional dispersion per galaxy. We note that the standard deviations of the bar length measurements are similar in each connection category ($\sim 17\%$) and unbiased.

\begin{table}
\begin{center}
  \begin{tabular}{r r r r} 
Connection & Ngal & $\langle \Delta L_g \rangle\%$ & $\sigma (\Delta L_g) \%$ \\ \hline
S (spiral to bar) & $433$ &$-0.5$ & $17.6$  \\
OR (only ring) & $219$ & $1.0$ & $16.7$  \\
NSR (no spiral or ring) & $61$ &  $-2.3$ & $11.1$ \\
R (sprial to ring to bar) & $49$ & $0.7$ & $11.8$ \\
SR (mixture of R) & $9$ & $8.8$ & $27.5$ \\ \hline
  \end{tabular}
\caption{\label{conn_res_cl} The bar, ring and spiral arm connection categories for the cleaned sample of galaxies with high classification agreement.  We show the number of galaxies in the sample Ngal, and the average and standard deviation of the fractional bar length scatter measurement.}
\end{center}
\end{table}

First, we note that the connection with the bar connecting directly to the spiral arms (denoted by S) is the most common (in $56.1\%$ of these galaxies). This result has been seen previously by earlier observations of $147$ galaxies \citep[][]{Buta:2005ym}. We extend this earlier work by using five times more barred galaxies. 

\begin{figure*}
   \centering
       \includegraphics[scale=0.4]{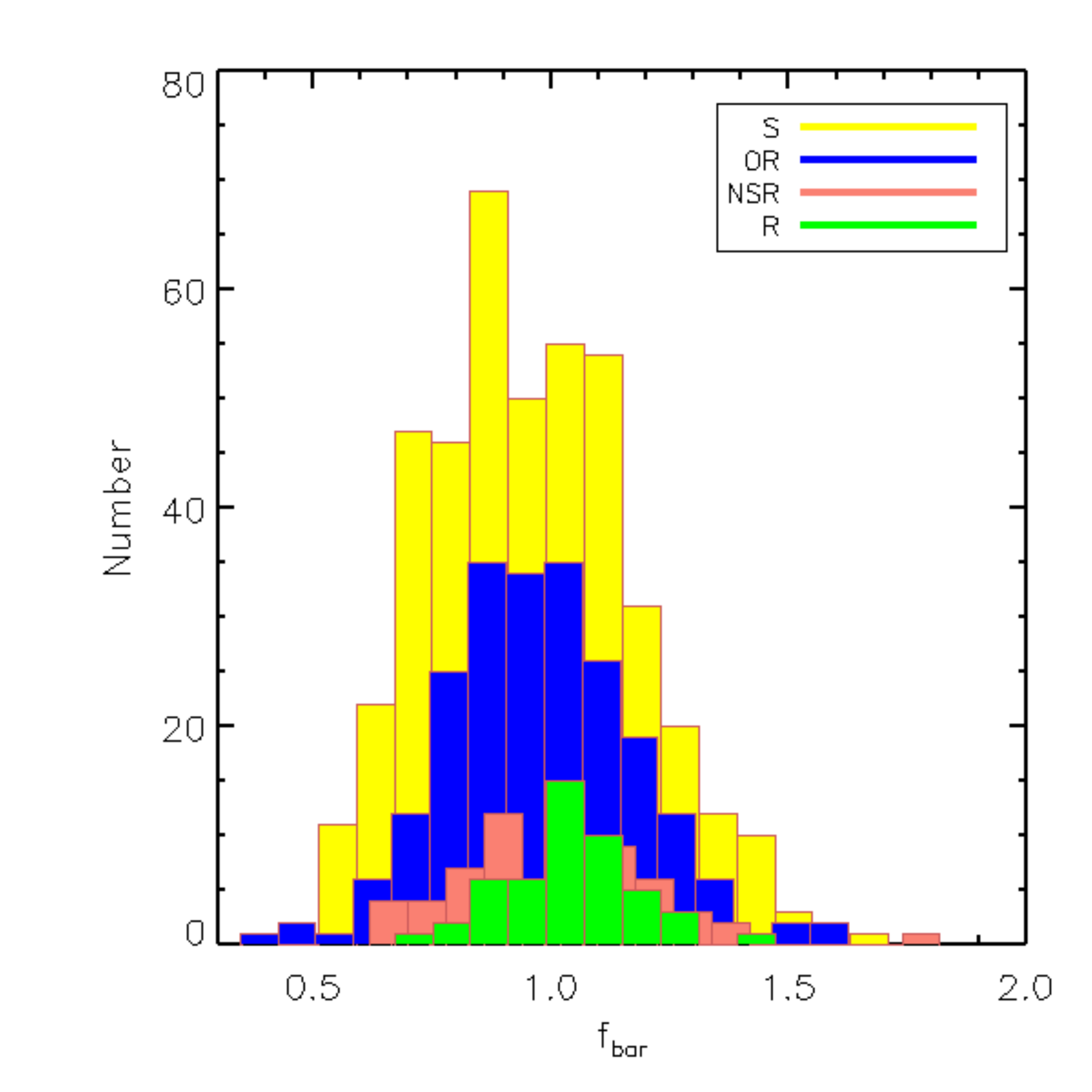}
            \includegraphics[scale=0.4]{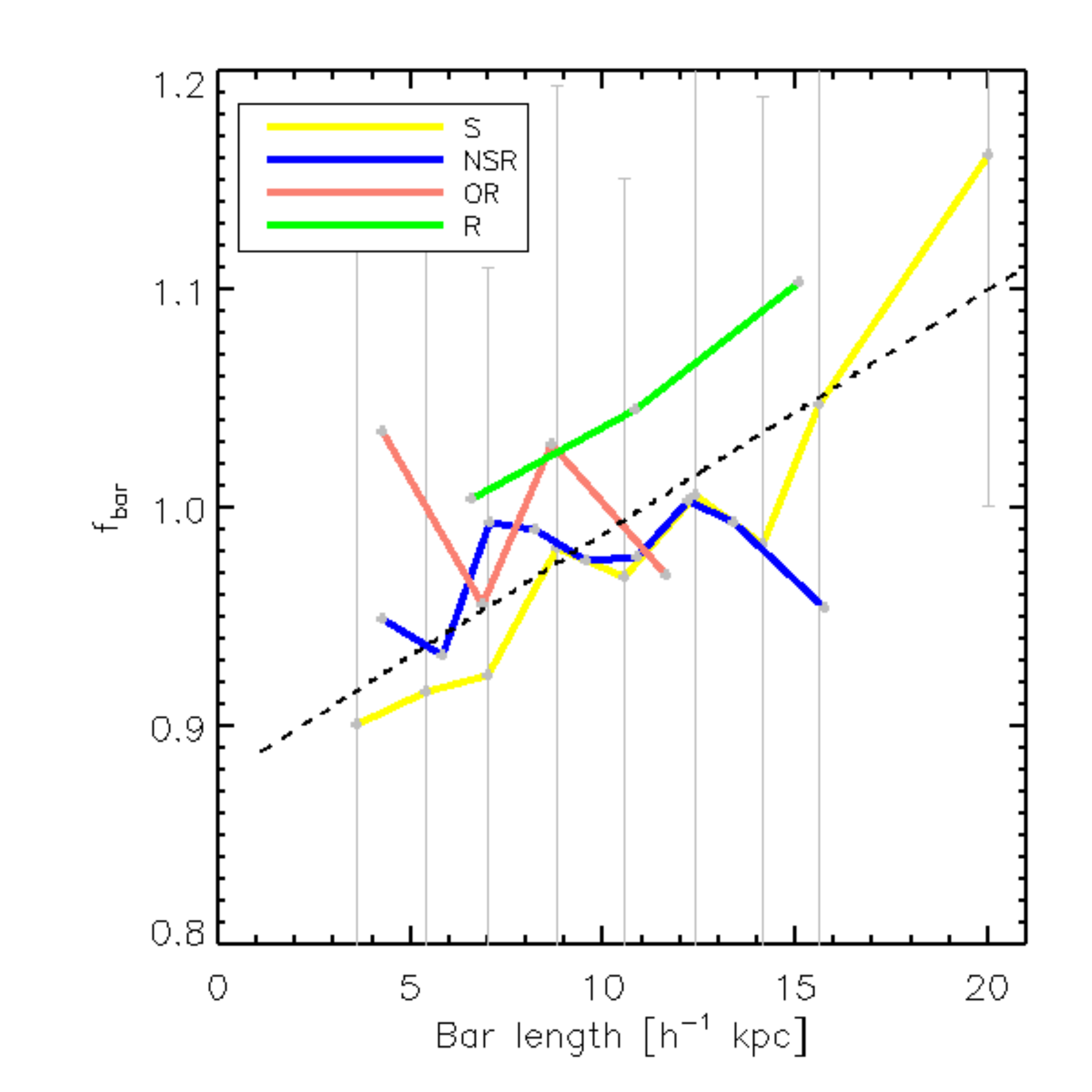}
   \caption{   \label{sprial_bar_connect1} We show the distributions of bar ellipticities $f_{bar}$ (left panel) and the correlation between $f_{bar}$ and bar length (right panel), for subsamples of galaxies split by how the spiral arms (if they exist) are connected to the ring (if it exists) and the galaxy bar. In the right panel the error bars show the standard deviation of the binned data for the connection types S, and the dashed line is the line of best fit to all the bar-spiral arm configurations.}
\end{figure*}

In Fig. \ref{sprial_bar_connect1}, we show how the distribution of bar ellipticities $f_{bar}$  (left panel) and the correlations between $f_{bar}$ and bar length (right panel) change with how the spiral arms (if they exist) are connected to the end of the bars or ring. We only show galaxy subsamples with more than $10$ galaxies (i.e. exclude the SR subsample), and the error bars show the standard deviation of the binned data for the connection types S. To calculate the bar strength $f_{bar}$, we follow \cite{2002MNRAS.336.1281W} and define
\begin{eqnarray}
f_{bar} = \frac{2}{\pi}\big( \arctan((b/a)_{bar})^{-0.5} - \arctan((b/a)_{bar})^{+0.5}\big) \nonumber
\end{eqnarray}
where $(b/a)_{bar}$ is the ratio of the bar width and bar length. $f_{bar}$ has been shown to be a proxy for bar strength \citep[][]{2007MNRAS.381..401L}. In the right panel of Fig. \ref{sprial_bar_connect1}, we show the correlation between bar strength $f_{bar}$ and length $L$ for different connections. The standard deviation of the S connection is shown using the grey error bars, we overplot the line of best fit for the combined samples, which is given by $f_{bar}=0.88+0.01\times L$ with a (un)reduced chi-squared of ($32.30$)$0.04$.

We find that the $f_{bar}$ medians of the S ($0.74 \pm 0.21$ with $1\sigma$ standard deviation) and OR ($0.77
\pm 0.20$) samples to be smaller than those of NSR ($1.17 \pm 0.21$) and  R ($1.14 \pm 0.14$). This implies that in  configurations where the bar is connected directly to the spiral arms, or where there are no spiral arms and the bar is connect to a ring, the bars are less elliptical (or are weaker) than those configurations where there are no spiral arms or rings, or the bar is connect to the ring and the ring is connected to the spiral arms. In the left panel of Fig.  \ref{sprial_bar_connect1} we see that the  bar strength increase for all bar-spiral arm configurations as a function of increases bar length, but that there is a large dispersion.  We note that the R connection has typically larger values of $f_{bar}$ (i.e., are stronger) than the other configurations, which is in agreement simulations \citep{2004ARA&A..42..603K,1980ApJ...235..803S}, but in contrast with \citep{Buta:2005ym} who find that galaxies with strong bars are more likely to have spiral arms, and that weaker bars are more likely to be ringed, when compared with the global average. 

We remind the reader that our barred galaxy sample in this section has been heavily cut to include only galaxies whose connection classification agreement is very high, but still contains five times more galaxies than any previous study.

\begin{figure*}
   \centering
   \includegraphics[scale=0.4]{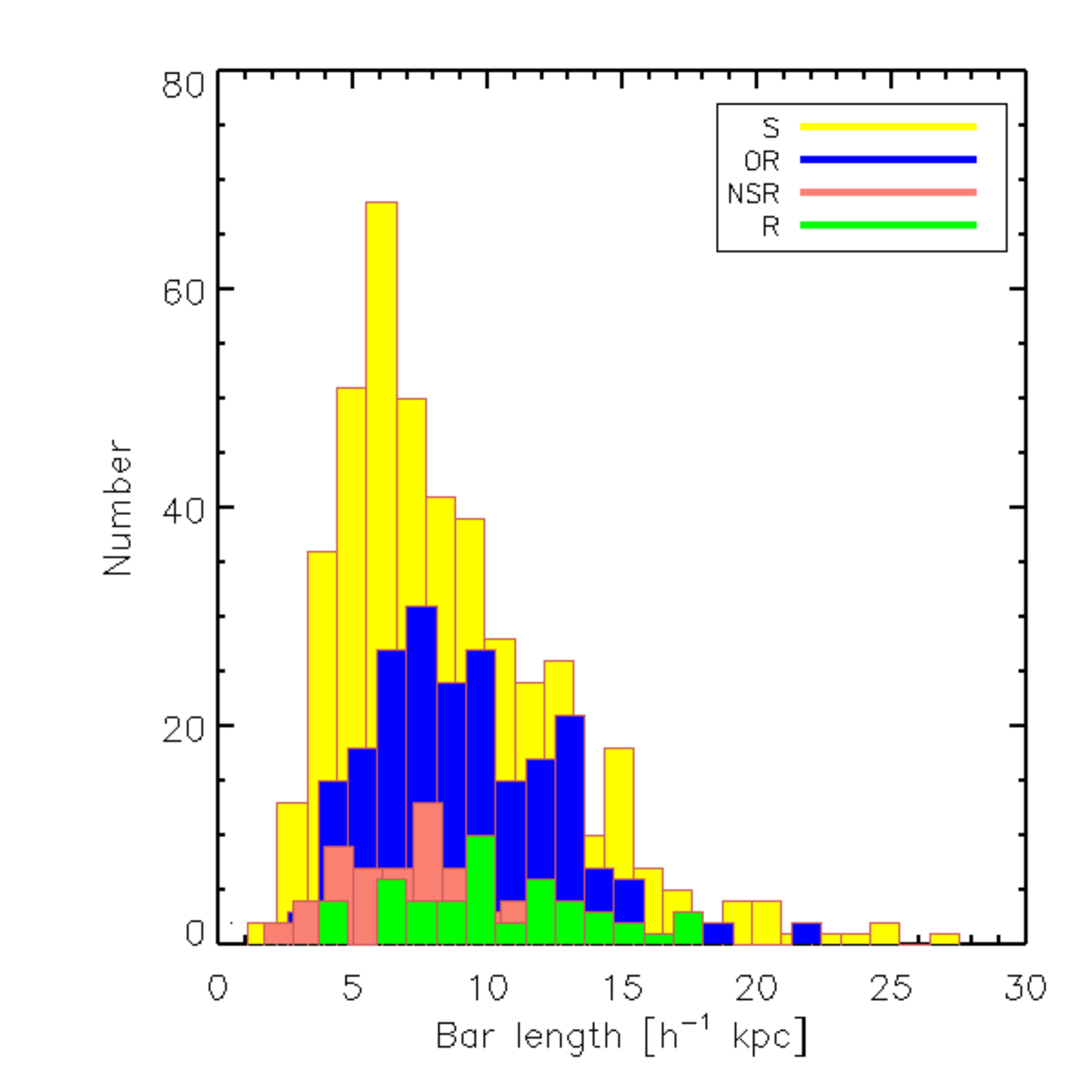} 
   \caption{   \label{sprial_bar_connect2} We show distributions of bar length (left panel) and galaxy color (right panel) for subsamples of galaxies split by how the spiral arms (if they exist) are connected to the ring (if it exists) and the galaxy bar. }
\end{figure*}

In Fig. \ref{sprial_bar_connect2}, we further explore the above results, by presenting histogram distributions showing how the subsample populations are distributed in bar length and galaxy color. In the left panel, we find that the distribution of bar lengths are similar from small to intermediate bar lengths ($<15\,h^{-1}$kpc), but the S connection can host longer bars than the other connections (which may be due to the larger sample size).  The median bar lengths (and $1\sigma$ standard deviation) of the different connections are given by; S: $7.44 \pm4.19 \,h^{-1}\,\kpc$, OR: $8.81 \pm3.48 \,h^{-1}\,\kpc$, NSR: $7.26 \pm2.72\, h^{-1}\,\kpc$, R: $9.94\pm3.61\, h^{-1}\,\kpc$.

\begin{figure*}
   \centering
   \includegraphics[scale=0.4]{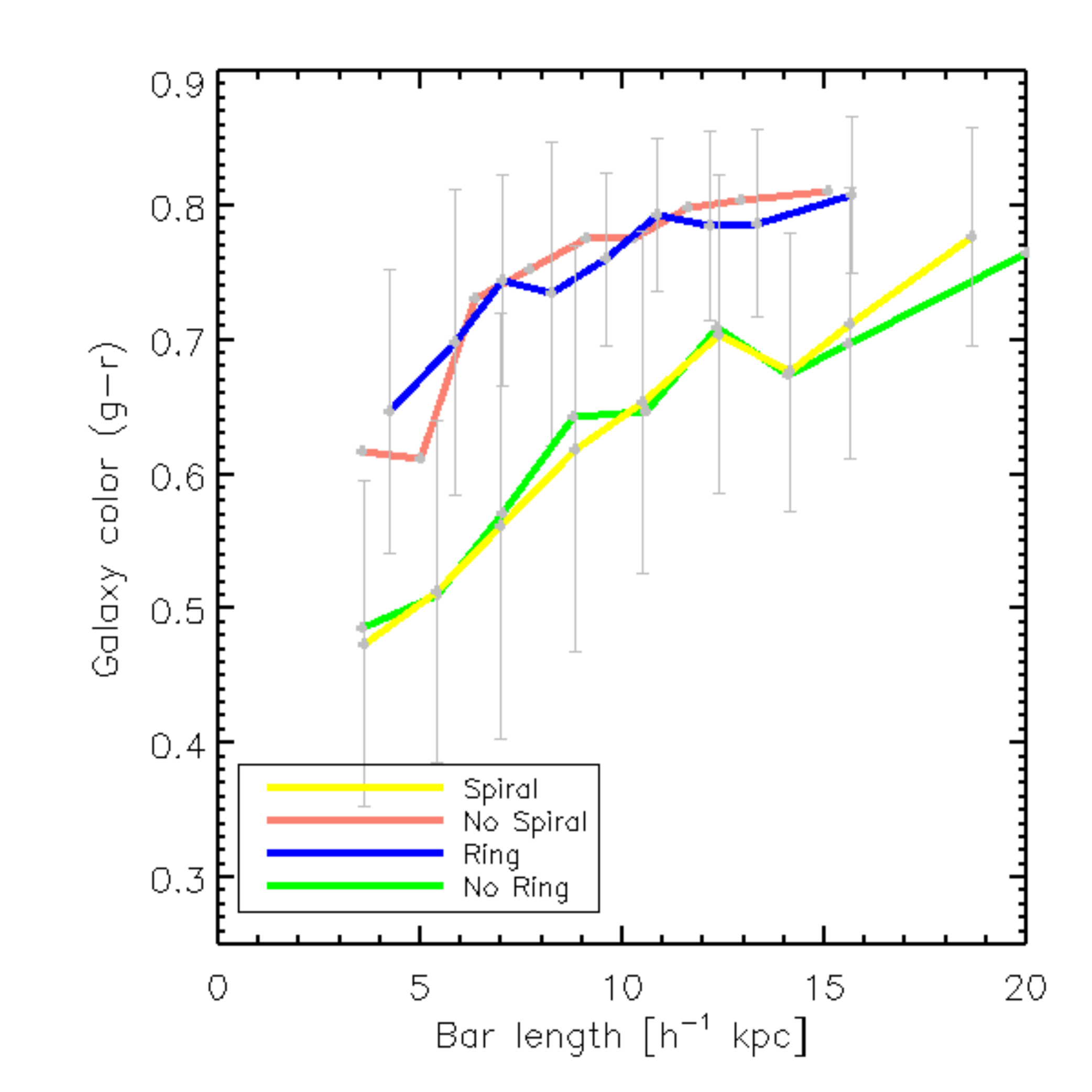}
       \includegraphics[scale=0.4]{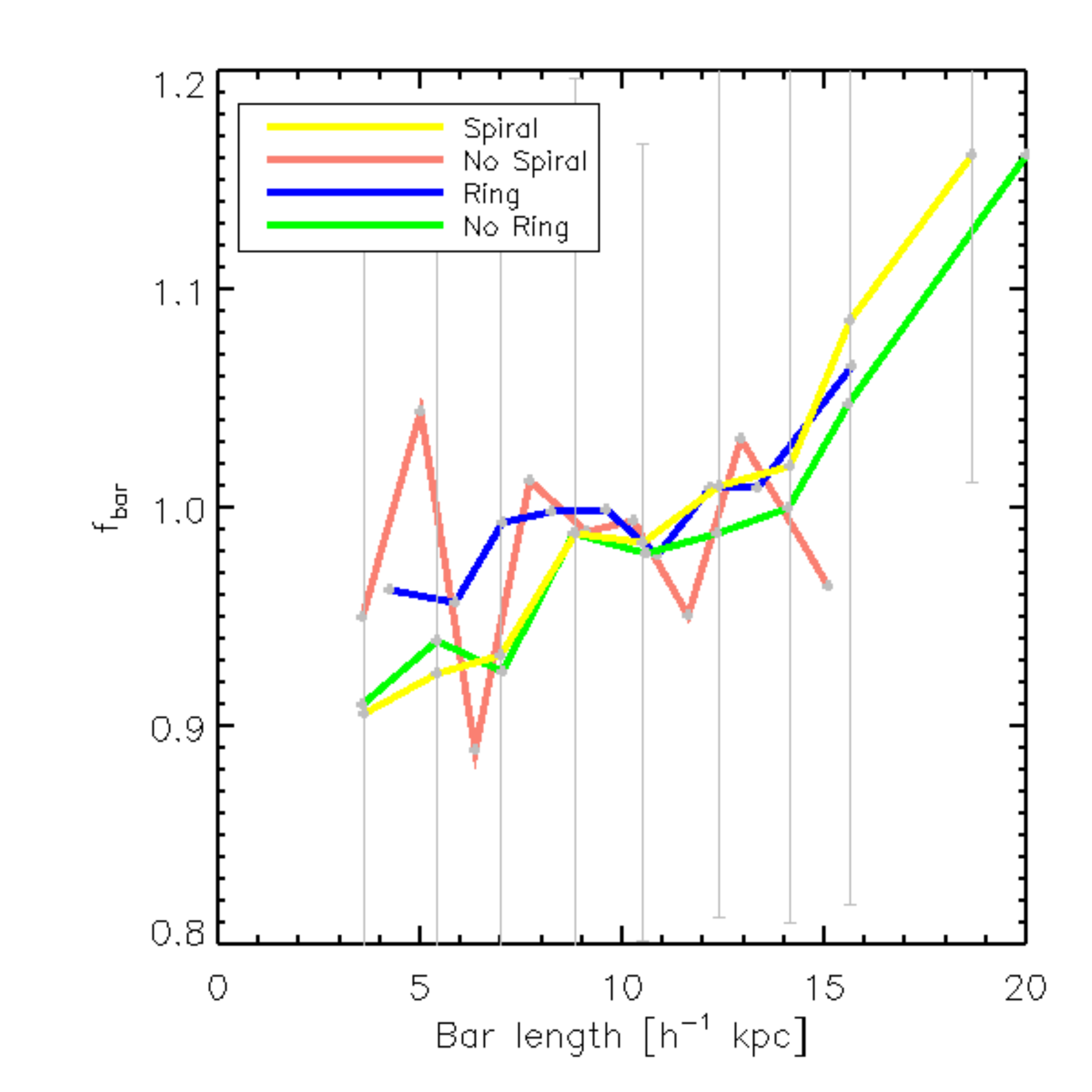}
   \caption{   \label{sprial_bar_connect3} We show bar length against galaxy color (right panel) and bar strength $f_{bar}$ (left panel), for grouped samples of galaxies split by existence, or not, of spiral arms and a ring. For ease of viewing, we show the error bars for selected connection types, which represent the standard deviation of the binned data.}
\end{figure*}

We note that the NSR (no spiral arms or ring are present) galaxies host bars which are typically the shortest (although the dispersion is moderate) and that the number of longer bars drops quickly, and there are no bars greater than $15\,h^{-1}$kpc.   The color histogram in the right panel of Fig. \ref{sprial_bar_connect2} shows the clear excess of blue S galaxies.  We also find that all the other connection subsamples have similar distributions, showing little differences in the bar length and galaxy color distributions, but we note (in passing) that all subsamples follow the same global trend, i.e. smaller bars are hosted in bluer galaxies, and galaxies become redder with increasing bar length (as per Fig. \ref{barcolbarlength} and the left panel in Fig. \ref{sprial_bar_connect3}).

We also group the galaxies into four combined categories, those with and without spiral arms (S and R and SR connections, totaling $491$ galaxies) and (OR and NSR, totaling $280$ galaxies), and those with and without a ring (R and OR and SR, totaling $277$ galaxies) and (S and NSR, totaling $494$ galaxies). In Fig. \ref{sprial_bar_connect3} we show galaxy color against bar length, and $f_{bar}$ (as a measure of bar strength) against bar length.
In the right panel of Fig. \ref{sprial_bar_connect3} we see that galaxies which host spiral arms or fail to host a ring, are bluer than those galaxies which host a ring, or fail to host spiral arms. The same increasing bar length with increasing redness trends are seen in all the samples, as before. In the left panel of Fig. \ref{sprial_bar_connect3} we see that bar length and $f_{bar}$ are independent of the presence or lack thereof, of spiral arms and rings with these groupings of data suggesting that the stronger bars seen in the R connection require a ring only (and no spiral arms).

\subsubsection{Full sample}
As an alternative to making strict cuts on the full galaxy sample to obtain galaxy subsamples, we can use all of the connection classifications for all galaxies. We do this by scaling the total number of votes per category per galaxy by the total number of classifications per galaxy. We need to do this because some galaxies have been classified more times than others (see \S\ref{website}). The total numbers of unscaled votes per category, and the corresponding number of scaled votes are shown in Table \ref{conn_res}. 

\begin{figure*}
   \centering
    \includegraphics[scale=0.4]{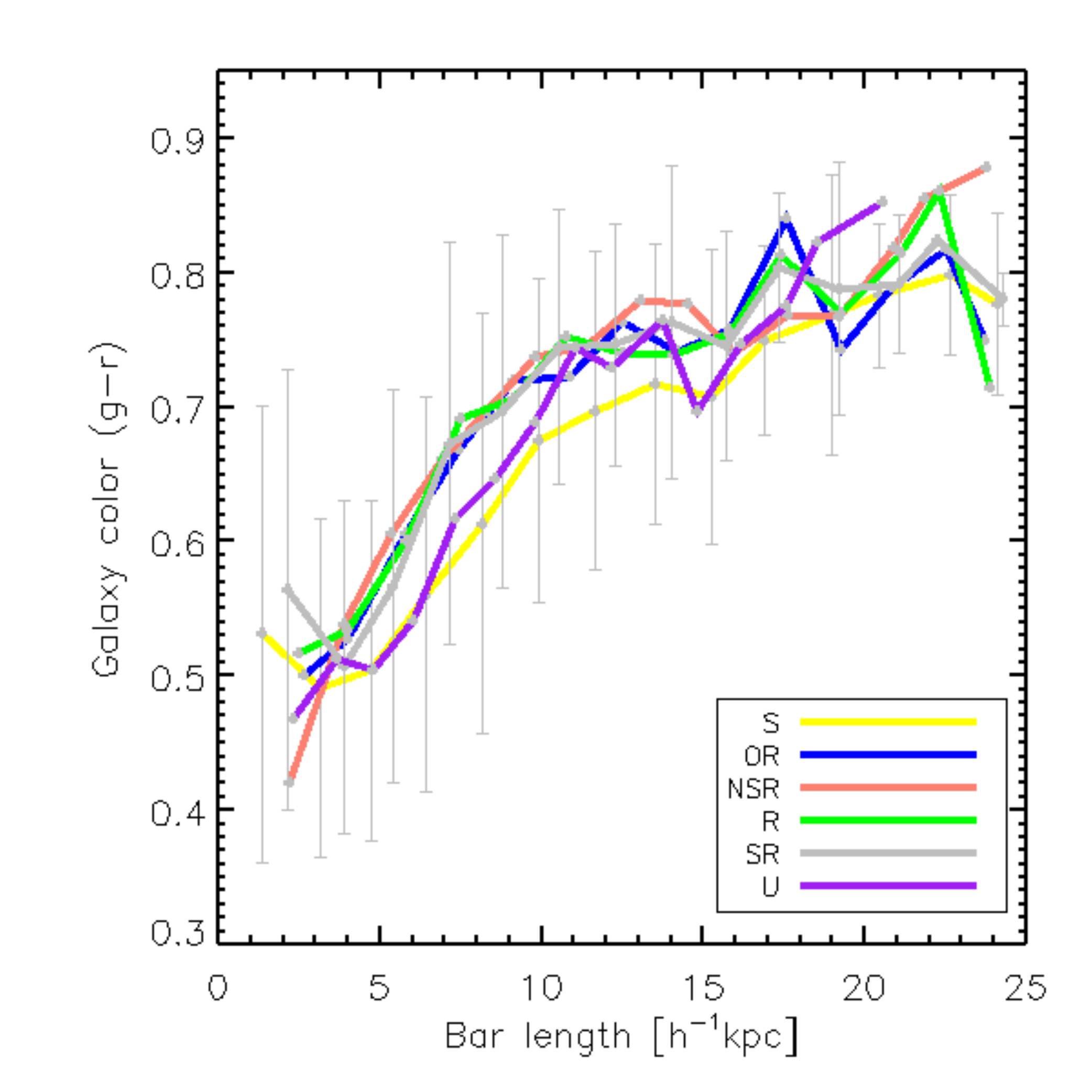} 
  \includegraphics[scale=0.4]{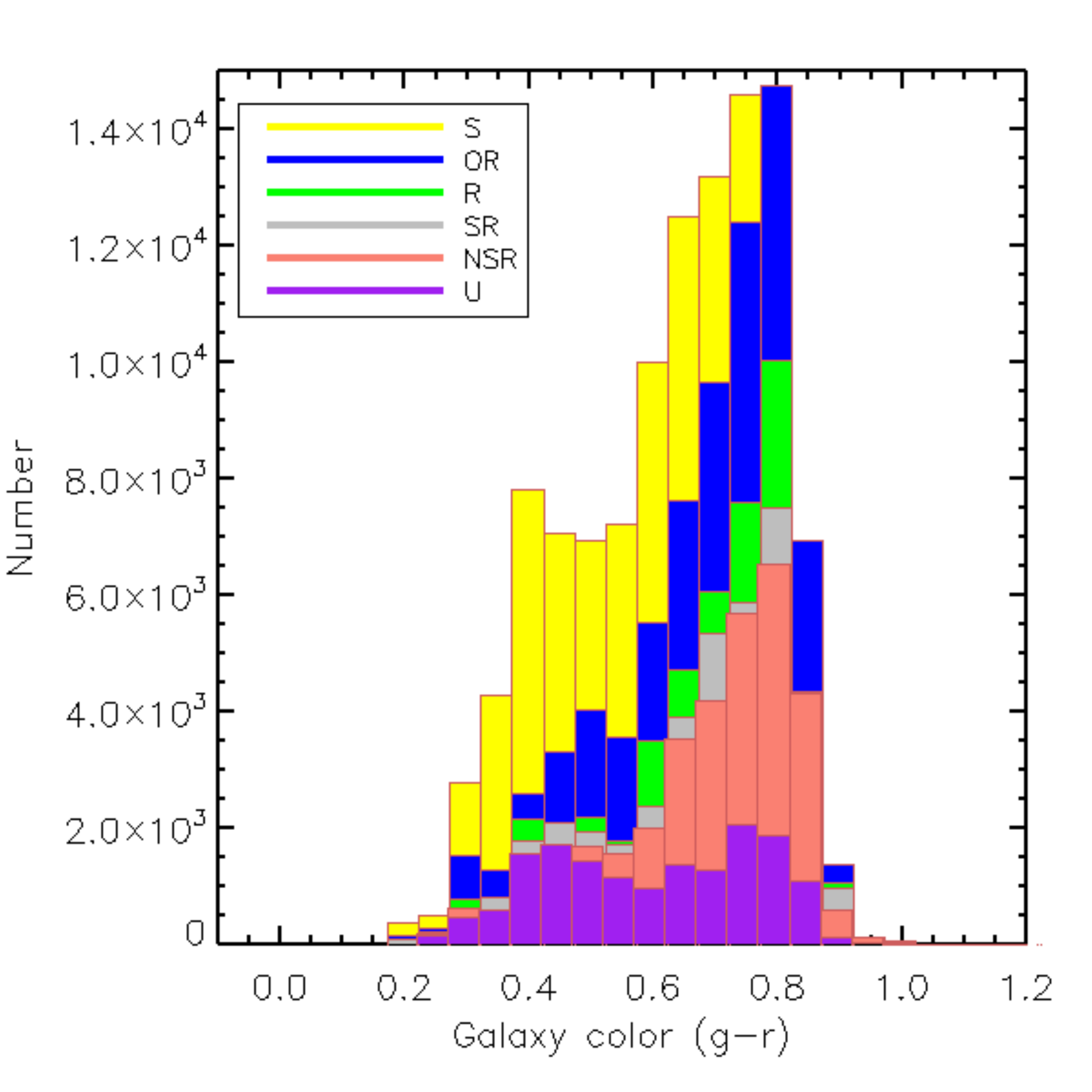}
   \caption{   \label{sprial_bar_connect4} We show bar properties for subsamples of galaxies split by how the spiral arms (if  they exist) are connected to the ring (if it exists) and the galaxy bar. The left panel shows the bar length against galaxy color for the full subsamples; the error bars show the standard deviation of the binned data for selected connection types. The right panel shows the galaxy color distributions.}
\end{figure*}

\begin{table}
\begin{center}
  \begin{tabular}{c r r} 
Connection & Total number & Scaled number    \\ \hline
S & $4334$ & $1032.23$ \\  
OR & $3294$ & $753.70$ \\ 
NSR & $1361$ & $339.08$ \\ 
R & $ 2208$ & $475.17$ \\ 
SR & $1774$ & $390.18$  \\ 
U & $631$ & $159.61$  \\  \hline
  \end{tabular}
\caption{\label{conn_res} The total number of bar and spiral arm connection classifications for all galaxies, and the scaled number of classifications, which allows galaxies with different number of classifications to be equally compared.}
\end{center}
\end{table}
First, we note that the S  galaxy population still contains more galaxies than the other connections, but the significance has been reduced to $32.77\%$ of the total number of votes. We are now able to view trends of all the connection populations, and show the bar length and galaxy color, and the galaxy color distribution, in Fig. \ref{sprial_bar_connect4}.

We find that the scaled subsamples are more similar to each other than the clean subsamples (in \S\ref{cleangals}), and that using the full sample removes most of the signal, as expected because a larger number of less obviously identifiable galaxies are included. The distributions of bar lengths and galaxy colors are also similar, and extend across the full range of bar lengths, as seen in the left panel of Fig. \ref{sprial_bar_connect4}. In each subsample we still see the same trend as before, that smaller bars are hosted in bluer galaxies, and bar length increases as galaxies become redder. The right panel of Fig. \ref{sprial_bar_connect4} shows the distributions of galaxy color. We still find an excess of bluer S galaxies, although it is less pronounced, and note the peak and distributions of the other subsamples are similar. The U (unsure or unable to make a classificaition) subsample shows a bimodal color distribution, one peak is located in the same location as the other subsamples, but the other peak is even bluer than the S connection, but we note that the scaled number of galaxies in this sample is a factor of ten smaller than the S subsample.

\section{conclusions and discussion}
\label{conclus}
We presented galaxy images selected to contain galactic bars (from Galaxy Zoo 2) to members of the Galaxy Zoo community using the Google Maps interface. The lengths and widths of $3150$ galactic bars were measured three or more times per galaxy, independent of previous measurements, and information describing how the galactic bars and spiral arms are connected, were collected.

We have shown that the sample of barred galaxies recovered by the GZ2 observers is unbiased, and thus our bar measurements are robust against systematic effects. We find that observers are able to reproduce their own bar length measurements to $ 0.5\pm 6\%$, and each others' bar length measurements to $10\pm14\%$. 

We now return to the questions posed in the introduction based on simulations and other data;
\begin{itemize}
\item{How do galaxy colors change as a function of bar length?}
\end{itemize}
We find a split in the color ($g-r$) and absolute $r$ band magnitude relation of our barred galaxy sample, described by redder early-type disk galaxies and their bluer, late-type disk galaxy counterparts. Remarkably, we can reproduce these populations by cutting on galactic bar length; longer bars ($>6\,h^{-1}$kpc) are found in early-type disk galaxies, and shorter bars, in late-type disk galaxies. We find the longest bars exist in the reddest disk galaxies, and that the shortest bars are found in the bluest disk galaxies. These findings are in agreement with recent bar studies of $253$ galaxies in the Virgo cluster \citep{Giordano:2010gg}, but in disagreement with simulations by \cite{Scannapieco:2010nw}.
\begin{itemize}
\item{How are the colors of galaxies, bars and disks correlated? }
\end{itemize}
Using the bar length measurements, we can estimate bar ($0<R<R_{bar}$) and disk ($R_{bar}<R<2\,R_{Petro}$) colors,  
and we show that these colors are correlated, in agreement with simulations by  \cite{Scannapieco:2010nw}, who find correlated bar and disk colors in a sample of eight galaxies. 
\begin{itemize}
\item{Are other galaxy properties affected by bar length?}
\end{itemize}
We also find that bar length is a function of galactic bulge size, in agreement with previous observational studies using $32$ galaxies \citep[][]{G1980A} and more recently with $300$ galaxies \citep[][]{Gadotti:2010ct}, and early simulations \citep[][]{Athanassoula:2003wd}, which suggest that bars become longer as they slow down and exchange angular momentum with the galactic bulge. We are able to expand, and further test this work, with measurements of $3150$ barred galaxies, and find that bar length divided by galaxy size increases linearly with bulge prominence with a gradient of $0.12\pm0.05$, i.e. the bar extends $12\%$ further into the disk in galaxies with a large bulge, compared to those without a central bulge. We do however find a large scatter in the bulge to bar relation (see Fig. \ref{fracdevbarlength}).
\begin{itemize}
\item{Is bar strength or length correlated with the presence of a ring?}
\end{itemize}
To examine this, we first build subsamples of galaxies split by the question ``How do the spiral arms (if they exist) connect to the ring (if it exists) and bar?". We only include galaxies whose classifications agree to high accuracy ($>80\%$). We then combine the subsamples in those galaxies which do (not) host spiral arms and those galaxies which do (not) host a ring. We compare with observations of $147$ galaxies \citep[][]{Buta:2005ym}, which suggest that galaxies with stronger bars have spiral arms and that weaker bars are more likely to be ringed, when compared with the global average, and simulations \citep{2004ARA&A..42..603K,1980ApJ...235..803S} which find the contrary. We use $f_{bar}$ \citep[][]{2002MNRAS.336.1281W}  as a proxy for bar strength \citep[][]{2007MNRAS.381..401L} and find that our samples agree with the simulations using a cleaned data sample of $771$ galaxies, and on the full sample (see the text below). However stronger bars are observed in galaxies which host a ring only if there are no spiral arms present.
\begin{itemize}
\item{How is the bar-to-spiral arm connection different in galaxies with longer bars?}
\end{itemize}
By selecting galaxies whose bars are directly connected to the spiral arms (connection S), we identify a sample of galaxies which are bluer than the other subsamples. This means that galaxies which host a ring, or fail to host spiral arms are typically redder than those which only host spiral arms.  In each of the subsamples we continue to identify shorter bars in bluer ($g-r$) galaxies, and see that bar length increases as galaxy become redder. We re-examine these trends by scaling the number of classifications per category per galaxy by the total number of classifications per galaxy. This allows the full $3150$ barred galaxy sample to be used statistically. We find the same trends as before, but at a lower significance.

Furthermore we find that in $56\%$ ($36\%$) of barred galaxies in the cleaned galaxy sample (in the full sample), the spiral arms are directly attached to the end of the bar. The preference for this configuration has been prediction from simulations  \citep[][]{Athanassoula:2009cw}, and confirmed by earlier observations of $147$ galaxies \citep[][]{Buta:2005ym}.
$\,$\\

We have demonstrated above, how the properties of simulated barred
disk galaxies agree extremely well with observed galaxies in general.
There are just a small number of ways in which current simulations do
not match our observations. For example, comparing bar length and galaxy color \citep[c.f. simulations by][]{Scannapieco:2010nw}.

$\,$\\
  
To aid further research and collaboration between simulators and observers to examine the above discrepancies, we are making the bar length and width measurements public, and can be found here: {\tt http://icc.ub.edu/$\sim$hoyleb/} (and also at {\tt http://data.galaxyzoo.org}). The comma separated file contains the SDSS unique object identifier (Objid), the average bar length and width measurements, their corresponding standard deviations, and the number of bar length measurements per galaxy.

\section*{Acknowledgments} 
\label{ack}
The authors would like to thank an anonymous referee for comments which improve the paper. During this work BH was partially funded by a grant from Google and grant number FP7-PEOPLE- 2007- 4-3-IRG n 20218, and the authors thank Google for their support of this and other Galaxy Zoo related projects. KLM acknowledges funding from the Peter and Patricia Gruber
Foundation as the 2008 Peter and Patricia Gruber Foundation IAU
Fellow; from a 2010 Leverhulme Trust Early Career Fellowship and from
the University of Portsmouth and SEPnet (www.sepnet.ac.uk). Support for the work of KS was provided by NASA through Einstein Postdoctoral Fellowship grant number PF9-00069 issued by the Chandra X-ray Observatory
Center, which is operated by the Smithsonian Astrophysical Observatory for and on behalf of NASA under contract NAS8-03060. The development of Galaxy Zoo 2 was supported by The Leverhulme Trust. Funding
for the SDSS and SDSS-II has been provided by the Alfred
P. Sloan Foundation, the Participating Institutions, the
National Science Foundation, the U.S. Department of
Energy, the National Aeronautics and Space Administration,
the Japanese Monbukagakusho, the Max Planck
Society, and the Higher Education Funding Council for
England. The SDSS Web Site is http://www.sdss.org/. 

\begin{widetext}
More specifically, this work would have been impossible without the help of the following GZ2 volunteers: Aaron Wilson, Adam Allford, Aileen Waite, Alan Eggleston, Alan L. Harris, Alberto Conti, Aldin Omanovic, Alice Sheppard, Allan Meadows, Amanda Peters, Andras Balogh, Andreas Lehnen, Andrew Crossett, Angeline Farrow, Anik Wellhausen, Arfon Smith, Augusta, BEC, BasAlbers, Brad Neuhauser, Brian Portlock, Briana Harder, Bruno Chiaranti, Bruno Domingues, Cacie Hart, Caro, Cathy Bialas, Channing Dutton, Charles Mulvey, Charlie Goodliffe, Chris Bogers, Chris H, Christopher Carrillo, Clive Higginson, Cope Middle School, Courtney Mitchell, Craig J Miller, Craig R. Sadler, David Burt, David Cortesi, David J Jones, Derek van Lessen, Dion Timmermann, Ed Buttery, Eduard Asensio, Edward Lilley, Eeva Saviranta, Elisabeth Baeten, Elizabeth Siegel, Elspeth Mitchell, Eric, Eric FABRIGAT, ErroneousBee, Essex Edwards, Fil, Frank Helk, Frank Helk, Gary Allen, Geoff Roynon, Gerwin Kappert, Graham Mitchell, Graham Robertson, Gravitroid, Greg Lang, HSDonnelly, Hank, Hannah Hutchins, Hannah Steiner, Hanny van Arkel, HansvdMfromTheNetherlands, Horsetuna, I.R. van de Stadt, Ian Rogers, Ingo Peters, Janice Kay McCollough, Jaureguiberry Alain, Jennifer Broekman, Jennifer Hicks, Jim Rajabally, Jo Syan Pitkethly, John Chunko, John Venables, John Zak, Jordan Bossard, Josh5555, Julia Wilkinson, JuniorUK, KY Receveur, Keith Hughitt, Krzysztof Olejnik, L, Briggs, Lanny Ripple, Lee Reiswig, Jr., Leila Sattary, Lily Lau WW, Logan Zoel, Lukas, Mairi Yates, Marc Lluell, Marcin Janus, Mark Gill, Mark Sands, Markku Autio, Markus Pioch, Martin McCarthy, Massimo Mezzoprete, Matt, Matt Swint, Matthew Holger, Mauro Marussi, Michael "Amberwolf" Elliott, Michael Josephs, Michael Josephs, Michael Wehner, Michael Willems, Michal Winiarski, Mihnea-Costin Grigore, Mike Chelen, Mike Need, Mike Tindle, Mike W T, Monica Butcher, NGC3172, Nancy Weitz, Nanne Sierkstra, Nils Weinander, Paola Riccucci, Patrick Logiste, Patrik AlzŽn, Paul D. Stewart, Peter Jennings, Peter Nixon, Philip Barker, RandyC, RapidEye, Raquel Shida, Rehana Rodrigues, Ret, Richard Daniel Armour, Rick Lopez, Rinus Rekkers, Rita Tojeiro, Robert Holdsworth, Robert Hubbard, Robert Simpson, Robin O-Malley, Rune Orsval, Ruth Graham, Sam Lehman, Sean Liddelow, Sebastian Matuschka, Shawn Pyle, Soren Scott, Stefan Ziel, Stephanie Carlino, Steve Dewey, Steve Malone, Teresa Neves, Thomas Kelder, Tim Sorbera, Todd Barrow, Tomas Hult, Tomas Vorobjov, Travis Becker, Yvonne Broer-Bloemen, biazawa, c\_cld, carlo artemi, chris randolph, danceswithwords, eric jolley, frame125, jim porter, mauidavejr61, mike russell, pep armengol, robertinventor, steven c. haack, tobias collierus, and $69$ other volunteers.
\end{widetext}

\bibliographystyle{mn2e}
\bibliography{bars}

\end{document}